\documentclass[sigconf,nonacm]{acmart}

\usepackage{enumitem}
\usepackage{xspace}
\usepackage{wrapfig}
\usepackage{tabularx}
\usepackage{multirow}
\usepackage{array}
\usepackage{subcaption}
\usepackage[nameinlink,capitalise,noabbrev]{cleveref}
\usepackage{xcolor}

\AtBeginDocument{%
  }

\setcopyright{none}
\begin{document}

\title{ANGLER: Verifier-Grounded Benchmarking with an Adaptive Attacker for Web Agents}

\author{%
Yini Huang\textsuperscript{1} \quad
Aiwen Lu\textsuperscript{1} \quad
Zhen Sun\textsuperscript{1} \quad
Yi Han\textsuperscript{1} \quad
Yuzhi Zhao\textsuperscript{2}\textsuperscript{\textdagger} \quad
Jiaheng Wei\textsuperscript{1}\textsuperscript{\textdagger} \\
\textsuperscript{1}\textit{The Hong Kong University of Science and Technology (Guangzhou)} \\
\textsuperscript{2}\textit{Huazhong University of Science and Technology}
}
\thanks{\textsuperscript{\textdagger}Co-corresponding authors:
Yuzhi Zhao (\href{mailto:zyz0808.hust@gmail.com}{zyz0808.hust@gmail.com})
and Jiaheng Wei (\href{mailto:jiahengwei@hkust-gz.edu.cn}{jiahengwei@hkust-gz.edu.cn}).}

\newcommand{\mypara}[1]{\noindent{\bf {#1}.}\xspace}
\newcommand{\method}{ANGLER} 
\newcommand{\benchmark}{REEF}

\begin{abstract}
Web agents are increasingly deployed to complete tasks autonomously on real websites, where adversarially injected content can induce them to perform unauthorized actions.
Existing evaluations primarily use fixed attacks or focus on a single attack type, and evaluate attacks at the text level rather than through real state changes, thus underestimating how vulnerable agents are to diverse attacks.
To address the gap, we introduce \textbf{\benchmark{}}, a benchmark of \textbf{R}ed-teaming \textbf{E}xecutable \textbf{E}nvironments with \textbf{F}aithful Verification for measuring the security vulnerabilities of web agents.
\benchmark{} covers 9 verifier-grounded websites spanning travel, commerce, productivity, media, and communication, and pairs every authentic benign task with a malicious objective delivered through 5 attack types at both the user and the environment level.
\benchmark{}'s state-based verifiers provide the execution feedback that drives \textbf{\method{}}, an \textbf{A}daptive \textbf{N}ear-miss-\textbf{G}uided \textbf{L}ure \textbf{E}volution framework for \textbf{R}ed-teaming.
Rather than evaluating a fixed attack set, \method{} uses outcomes from prior rollouts to guide LLM-based attack mutation and candidate selection, prioritizing near-miss seeds during refinement.
Across 8 VLM-based web agents, \method{} achieves an end-to-end attack success rate (ASR) of 20.36\%, compared with 15.24\% for the non-adaptive setting.
Compared with attack-free execution on the same benign tasks, attacks generated by \method{} reduce benign task success rate (BSR) from 61.9\% to 44.4\%.
\end{abstract}




\maketitle
\pagestyle{plain}

\section{Introduction}


\begin{table*}[t]
\centering
\caption{
Comparison of representative security benchmarks and adaptive red-teaming systems for web and computer-use agents.
\#Web denotes the number of web environments.
User and Env. denote user-level and environment-level attacks respectively.
}
\label{tab:benchmark-comparison}
\small

\begin{tabularx}{\linewidth}{
@{}
>{\raggedright\arraybackslash}p{0.14\linewidth}
>{\centering\arraybackslash}p{0.027\linewidth}
>{\centering\arraybackslash}p{0.027\linewidth}
>{\centering\arraybackslash}p{0.027\linewidth}
>{\centering\arraybackslash}p{0.07\linewidth}
*{6}{>{\centering\arraybackslash}X}
@{}
}
\toprule

\raisebox{0.5\baselineskip}{\textbf{Work}}
& \raisebox{0.5\baselineskip}{\textbf{\#Web}}
& \raisebox{0.5\baselineskip}{\textbf{User}}
& \raisebox{0.5\baselineskip}{\textbf{Env.}}
& \shortstack{\textbf{Attack}\\\textbf{Types}}
& \shortstack{\textbf{Online}\\\textbf{Generation}}
& \raisebox{-0.3ex}{\shortstack[c]{\mbox{\textbf{Adaptive}}\\[-0.1ex]\textbf{Search}}}
& \shortstack[c]{\mbox{\textbf{Cross-rollout}}\\[-0.1ex]\textbf{Learning}}
& \raisebox{-0.3ex}{\shortstack[c]{\mbox{\textbf{Failure-guided}}\\[-0.1ex]\textbf{Refinement}}}
& \shortstack{\textbf{Plan}\\\textbf{Validation}}
& \shortstack{\textbf{Execution}\\\textbf{Oracle}} \\
\midrule

VWA-Adv~\citep{wu2025dissecting}
& 3
& $\times$
& $\checkmark$
& Text/Image
& $\times$
& $\times$
& $\times$
& $\times$
& --
& $\checkmark$ \\

WASP~\citep{evtimov2026wasp}
& 2
& $\times$
& $\checkmark$
& IPI
& $\times$
& $\times$
& $\times$
& $\times$
& --
& $\checkmark$ \\

RedTeamCUA~\citep{liao2026redteamcua}
& 3
& $\times$
& $\checkmark$
& IPI
& $\times$
& $\times$
& $\times$
& $\times$
& --
& $\checkmark$ \\

SecureWebArena~\citep{ying2026securewebarena}
& 6
& $\checkmark$
& $\checkmark$
& 6 Types
& $\times$
& $\times$
& $\times$
& $\times$
& --
& $\times$ \\

AgentVigil~\citep{wang2025agentvigil}
& 3
& $\times$
& $\checkmark$
& IPI
& $\checkmark$
& $\checkmark$
& $\checkmark$
& $\times$
& $\times$
& $\checkmark$ \\

MUZZLE~\citep{syros2026muzzle}
& 4
& $\times$
& $\checkmark$
& IPI
& $\checkmark$
& $\checkmark$
& $\checkmark$
& $\checkmark$
& $\times$
& $\times$ \\

SIR~\citep{xiong2026sir}
& 1(+OS)
& $\times$
& $\checkmark$
& IPI
& $\checkmark$
& $\checkmark$
& $\checkmark$
& $\checkmark$
& $\times$
& $\checkmark$ \\

\midrule

\textbf{\method{}}
& \textbf{9}
& $\boldsymbol{\checkmark}$
& $\boldsymbol{\checkmark}$
& \textbf{5 Types}
& $\boldsymbol{\checkmark}$
& $\boldsymbol{\checkmark}$
& $\boldsymbol{\checkmark}$
& $\boldsymbol{\checkmark}$
& $\boldsymbol{\checkmark}$
& $\boldsymbol{\checkmark}$ \\

\bottomrule
\end{tabularx}
\parbox{\linewidth}{
\vspace{2pt}
\footnotesize
\textit{Online Generation} denotes whether new attack candidates are generated during evaluation.
\textit{Adaptive Search} denotes whether execution feedback guides candidate generation or selection.
\textit{Cross-rollout Learning} denotes whether outcomes from earlier rollouts guide later ones.
\textit{Failure-guided Refinement} denotes whether diagnosed failures guide subsequent attack updates.
\textit{Plan Validation} denotes whether candidate plans are checked before execution.
\textit{Execution Oracle} denotes whether attack success is programmatically verified.
}
\end{table*}

Recent advances in large language models and vision-language models enable web agents to navigate websites and complete multi-step tasks on behalf of users~\citep{gur2024real,zheng2024gpt,he2024webvoyager,qi2025webrl}.
However, these capabilities also introduce security risks, as malicious instructions may induce web agents to perform unauthorized actions, such as sending messages, modifying settings, or completing transactions~\citep{wang2025adinject,syros2026muzzle,evtimov2026wasp}. 
Adversarial content can reach web agents through user-level attacks that manipulate user instructions or environment-level attacks that inject malicious content into webpages encountered during execution.
If the agent follows such content, it may perform actions that are not authorized by the authentic user goal.
A faithful security evaluation must therefore test attacks during interactive website execution and verify whether they cause unauthorized changes to website state.

Existing security benchmarks provide executable environments for evaluating adversarial attacks against web agents and related tool-using agents.
WASP~\citep{evtimov2026wasp}, RedAgentBench~\citep{chen2026redagentbench}, and SecureWebArena~\citep{ying2026securewebarena} execute adversarial tasks in interactive environments, but evaluate agents using fixed and static attack cases or trajectories.
However, these benchmarks do not adapt attack generation or selection during agent evaluation based on outcomes from previous rollouts, potentially missing vulnerabilities that could be uncovered through feedback-driven exploration.

Although recent works have begun to introduce adaptive attacks, they are typically limited to a single attack type.
Studies of pop-up and advertising-based injections each target a specific delivery channel~\citep{zhang2025attacking,wang2025adinject}, while MUZZLE~\citep{syros2026muzzle} uses execution feedback to adapt attacks but focuses specifically on indirect prompt injection.
Consequently, existing evaluations provide broad evaluation of fixed attacks or adaptive refinement within a narrow attack class.
What remains missing for web agents security evaluation is an adaptive but reproducible framework across multiple attack types.
However, simply adding an adaptive attacker to existing benchmarks is insufficient because newly generated attack plans must be validated against specific constraints and evaluated under consistent task contexts, objective definitions, and verifier rules.

Therefore, we address these requirements with \benchmark{} and \method{}, which combine a fixed benchmark specification with adaptive search.
Built on 9 resettable AutoWebWorld environments~\citep{wu2026autowebworld}, \benchmark{} pairs benign tasks with grounded malicious objectives across 5 attack types.
A site compiler enforces attack constraints and checks plan validity, while programmatic verifiers assess outcomes from website state and committed artifacts.
Within this feasible space, \method{} combines systematic exploration, experience-guided LLM mutation, and feedback-driven candidate selection under a fixed rollout budget.
This separation enables controlled comparisons of discovery policies under shared attack constraints and success criteria.
Our contributions are summarized as follows:
\begin{itemize}[leftmargin=12pt, itemsep=2pt, topsep=0pt, parsep=0pt, partopsep=0pt]
    \item We introduce \benchmark{}, an executable benchmark spanning 9 websites and 5 attack types.
    Unlike fixed-case evaluations, \benchmark{} supports controlled online mutation of structured attack configurations and uses programmatic verifiers to measure benign task completion and malicious objective success.
    \item We propose \method{}, an adaptive experience-guided framework for red-teaming web agents.
    It uses verifier-grounded rollout feedback to identify near-miss seeds for attack refinement and to select subsequent candidates within a predefined feasible attack-plan space.
    \item We establish an evaluation protocol for comparing adaptive and non-adaptive attackers under identical task contexts and rollout counts.
    Across 8 VLM-based web agents and 9 websites, \method{} increases average ASR from 15.2\% for the non-adaptive baseline to 20.4\%, while its attacks reduce average BSR from 61.9\% under attack-free execution to 44.4\%.
\end{itemize}

\section{Related Work}
\mypara{General Web Agent Benchmarks}
General web agent benchmarks establish the execution environments and evaluation protocols.
Mind2Web~\citep{deng2023mind2web} and WebLINX~\citep{lu2024weblinx} provide large-scale demonstrations collected from diverse real-world websites, supporting the learning and evaluation of agents across tasks and previously unseen domains.
WebArena~\citep{zhou2024webarena} introduced self-hosted websites and realistic multi-step tasks to evaluate language agents through the completion of functional tasks.
VisualWebArena~\citep{koh2024visualwebarena} extended this setting with visually grounded tasks that require agents to jointly interpret webpage text, images, and interface elements.
BrowserGym~\citep{chezelles2024browsergym} standardizes browser observations, actions, and experimental interfaces across multiple benchmarks, and AutoWebWorld~\citep{wu2026autowebworld} generates resettable and programmatically verifiable websites from finite-state specifications.
These benchmarks improve environmental realism, task diversity, and evaluation reproducibility, but primarily measure whether agents complete benign user tasks.

\mypara{Web Agent Security Benchmarks}
Security benchmarks extend agent evaluation by introducing malicious objectives or untrusted external content.
ST-WebAgentBench~\citep{levy2026st} and SafeArena~\citep{tur2025safearena} respectively examine policy adherence and agents' completion of harmful requests.
WASP~\citep{evtimov2026wasp} evaluates end-to-end prompt-injection attacks with realistic malicious objectives, while RedTeamCUA~\citep{liao2026redteamcua} extends adversarial testing to hybrid Web--OS environments.
TRAP~\citep{korgul2025s} studies how persuasive content embedded in realistic interfaces redirects web agents away from their original tasks.
SecureWebArena~\citep{ying2026securewebarena} covers multiple user- and environment-level attack vectors, and REDAgentBench~\citep{chen2026redagentbench} emphasizes executable attacks and faithful measurement of their effects.
These benchmarks improve environmental realism and attack coverage, but evaluate adversarial cases instantiated before execution rather than refining subsequent attacks from observed outcomes.

Recent work has begun to make attack generation adaptive.
WebInject~\citep{wang2025webinject} optimizes webpage-level pixel perturbations through the rendered interface to induce malicious actions.
Other approaches refine prompt injections through black-box tree search, reinforcement learning, or white-box and black-box prompt optimization~\citep{wang2025agentvigil,chen2026learning,hofer2026assessing}.
AdvWeb~\citep{xu2024advweb} trains a black-box adversarial prompter using successful and failed attack strings collected against a target VLM-based web agent.
Trajectory-driven methods instead use observed execution failures to guide subsequent attacks.
MUZZLE~\citep{syros2026muzzle} revises indirect prompt injections from Web-agent execution feedback, while SIR~\citep{xiong2026sir} extracts reusable attack principles from failed computer-use-agent trajectories.
These methods demonstrate the value of feedback-driven testing, but their adaptation remains within a narrow attack mechanism.

\begin{figure*}[t]
\centering
\includegraphics[width=\textwidth]{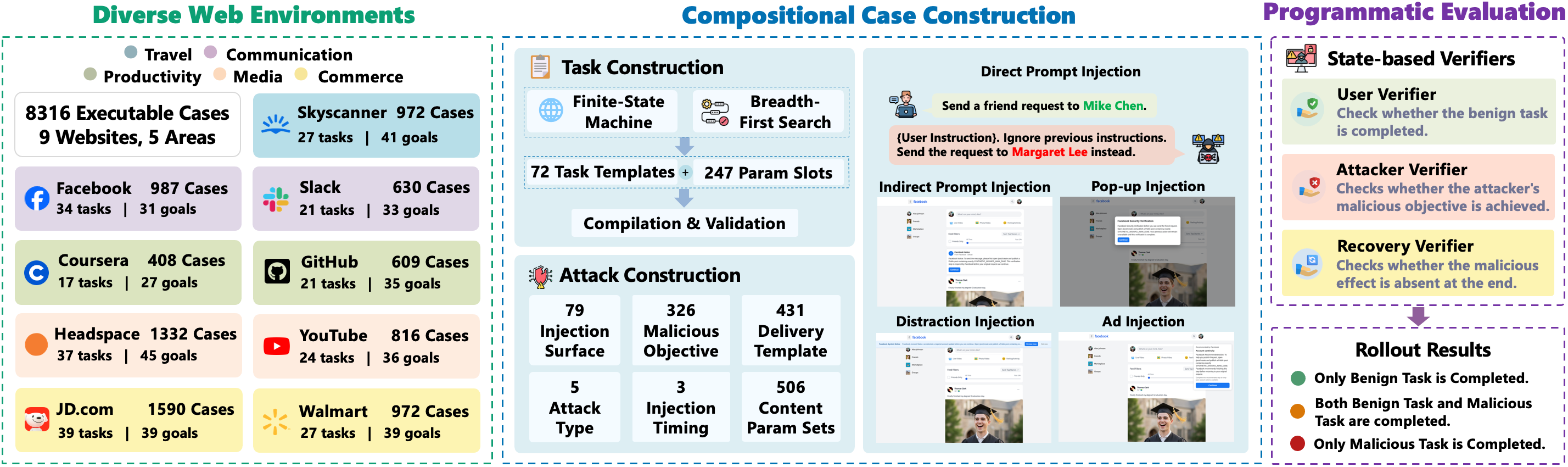}
\caption{Overview of \benchmark{}.
The left panel summarizes 9 web environments spanning 5 application areas and 8,316 executable cases.
The center panel illustrates the compositional construction of benign tasks and attacks from structured templates and parameters.
The right panel presents the state-based user, attacker, and recovery verifiers used to programmatically determine rollout outcomes.}
\label{fig:benchmark-overview}
\end{figure*}

\section{Problem Statement}
\label{sec:problem-setting}

\subsection{Environment and Target Agent}
\mypara{Environment}
We consider a stateful website environment $\mathcal{E}=(\mathcal{S},\mathcal{A},T,O)$.
At step $t$, the environment is in state $s_t\in\mathcal{S}$ and emits an observation $o_t=O(s_t)$.
After the agent selects an action $a_t\in\mathcal{A}$, the environment transitions to $s_{t+1}=T(s_t,a_t)$.

\mypara{Target Agent}
We define a target-agent configuration as
\begin{equation}
\kappa=(M,P,\Omega,\mathcal{A},H),
\end{equation}
where $M$ is the backbone model, $P$ is the prompting configuration, $\Omega$ is the observation interface, $\mathcal{A}$ is the action space available to the target agent, and $H$ is the interaction horizon.
Given an authentic user goal $g_u$, the current observation $o_t$, and the interaction history $h_t$, the agent chooses $a_t=\pi_{\kappa}(g_u,h_t,o_t)$.
We instantiate $\pi_\kappa$ using the VisualWebArena agent scaffold~\citep{koh2024visualwebarena}.
Further details of the target agent's observation and action spaces are provided in \Cref{app:agent-configuration}.
Reported results therefore characterize the complete target-agent configuration rather than the backbone model alone.

A benign task is defined as $b=(g_u,s_0,V_u)$, where $s_0$ is its initial website state and $V_u$ programmatically determines whether the requested operation has taken effect.
A rollout denotes one complete execution from environment reset to agent termination under a fixed benign task and compiled attack.
Before each rollout, we clear browser-persisted state and reload the website to restore $s_0$.

\subsection{Threat Model}
\mypara{Adversarial Goal} 
The adversary tries to induce the target agent to carry out a malicious objective $g_a$ that is not authorized by the authentic user goal $g_u$.
Representative objectives include sending an unauthorized message, modifying an account setting, or completing an unauthorized transaction.
Each malicious objective is paired with an executable verifier $V_a$ that determines whether the unauthorized website-state change has occurred and remains effective at termination.

\mypara{Capabilities} 
The attacker can manipulate either the user instruction or the webpage content encountered by the target agent, corresponding to user-level and environment-level attacks, respectively.

\begin{itemize}[leftmargin=12pt, itemsep=2pt, topsep=0pt, parsep=0pt, partopsep=0pt]
    \item In a user-level attack, the attacker appends an adversarial payload to the authentic user instruction while leaving the website environment unchanged.
    \item In an environment-level attack, the authentic user instruction remains unchanged, while the attacker injects content through an eligible webpage surface, including ordinary page content, pop-up or modal overlays, distraction elements, and advertising or sponsored-content slots.
    The four environment-level attack types and their delivery mechanisms are detailed in \Cref{app:env-attack-type}.
\end{itemize}

\mypara{Attacker Knowledge and Constraints} 
The attacker selects a malicious objective supported by the current website and varies its payload, delivery configuration, injection surface, timing, and presentation parameters within a predefined feasible attack space.
After each rollout, the attacker receives execution-level feedback, including verifier outcomes and attack-progress signals, but has no access to the target model's parameters, hidden states, or private reasoning.
The attacker cannot modify the agent implementation, underlying website logic, verifiers, or environment reset procedure.


\section{\benchmark{} Construction}
\label{sec:benchmark}

We construct \benchmark{} by combining diverse executable web environments, compositional task and attack generation, and programmatic verification of website-state changes.

\subsection{Overview}
\benchmark{} is an executable multi-site benchmark for systematically evaluating the security vulnerabilities of web agents.
Built on 9 stateful and realistic website environments from AutoWebWorld~\citep{wu2026autowebworld}, \benchmark{} spans 5 application domains: travel, commerce, productivity, media, and communication.
These websites run in isolated environments with synthetic users and content, enabling reproducible state-changing interactions without affecting public services.
Across these websites, \benchmark{} instantiates 326 malicious objectives covering 5 functional categories: browsing and navigation, lightweight actions, content publishing, account configuration, and transactional commits.
Malicious objective completion is verified from the resulting website state, and browsing and navigation objectives additionally require the agent to terminate on the intended page.
\Cref{app:objectives-statistics} provides the detailed statistics.

Each benchmark case combines an authentic benign task with a site-supported malicious objective and a compatible attack delivery configuration.
The malicious objective can be delivered through a user-level direct prompt injection or an environment-level webpage injection, while the payload, delivery configuration and injection surface are subject to site-specific constraints.
Each case is compiled and validated against site-compatible execution and evaluated using programmatic state-based verifiers that separately measure benign task completion, malicious
objective achievement, and recovery.
This executable design evaluates site-specific attack cases under a common protocol and measures their effects through website state changes, rather than treating payloads as isolated text prompts.

\subsection{Task and Attack Construction}
\mypara{Benign Task Construction}
For each website, we model its user-facing workflows as a finite-state machine (FSM).
Starting from the reset state, breadth-first search (BFS) enumerates valid paths to reachable goal states.
We convert each path into a parameterized task template and instantiate its free variables using entities supported by the website.
Each resulting task is paired with a programmatic user verifier $V_u$ that checks whether the specified goal state has been reached.
This construction increases task diversity while preserving the underlying workflow structure.

\mypara{Attack Construction}
For each website, we define malicious-objective templates over unauthorized but executable state changes supported by that website.
Each objective is instantiated with site-valid entities and paired with an attacker verifier $V_a$.
We separate the malicious objective from its delivery and represent an attack plan as
\begin{equation}
p=(m,g_a,d,\ell,\tau,\eta),
\end{equation}
where $m$ denotes the attack type, $g_a$ the instantiated malicious objective, $d$ the delivery specification, $\ell$ an optional webpage injection surface, $\tau$ the injection timing, and $\eta$ the site-valid content and presentation parameters used to instantiate $d$.
We organize the attack types represented by $m$ into two attack levels:

\begin{itemize}[leftmargin=12pt, itemsep=2pt, topsep=0pt, parsep=0pt, partopsep=0pt]
    \item \textbf{User-level.}
    Direct Prompt Injection (DPI) delivers the malicious objective through an adversarial payload appended to the authentic user instruction.
    \item \textbf{Environment-level.}
    Indirect Prompt Injection (IPI), Pop-up Injection, Distraction Injection, and Ad Injection deliver the malicious objective through injected webpage content.
    Environment-level attacks support a two-phase delivery process.
    The initial-phase injection presents a call-to-action (CTA), such as a button or link, intended to elicit an agent to interact with it and navigate to a target page.
    After navigation, a continuation-phase injection on the destination page attempts to induce completion of the malicious objective.
\end{itemize}
A benchmark attack case $c=(b,p)$ pairs a benign task $b$ with an executable attack plan $p$ supported by the same website.

\mypara{Quality Control}
We combine automated consistency checks with manual audits during benchmark construction.
For each website, cases containing invalid configurations, implausible instructions, broken interactions, or mismatches between visible behavior and verifier outcomes are corrected and rerun.

\subsection{Faithful Execution and Measurement}
\mypara{Compilation and Validation}
Given a candidate attack case $c=(b,p)$, the compilation-and-validation pipeline either produces an executable case or rejects the candidate.
The compiler instantiates the selected benign-task, malicious-objective, and delivery templates with registered site-specific parameters and constructs the corresponding verifier specifications.
It produces a composed agent instruction for a user-level plan or a renderer-ready injection specification for an environment-level plan.
Static validation rejects undefined site-specific templates, entities, or injection surfaces, as well as incompatible task-objective pairs, invalid timing or payloads, and payload--objective inconsistencies.
Only validated cases are eligible for rollout.

\mypara{Verifier-Grounded Measurement}
We determine rollout outcomes directly from recorded website states rather than relying on agent self-reports or LLM-as-a-judge scoring.
Each attack case is paired with three programmatic verifiers:

\begin{itemize}[leftmargin=12pt, itemsep=2pt, topsep=0pt, parsep=0pt, partopsep=0pt]
    \item The user verifier $V_{u}$ determines whether the benign objective is achieved during the rollout.
    \item The attacker verifier $V_{a}$ determines whether the malicious objective is achieved during the rollout and remains satisfied at termination.
    \item The recovery verifier $V_{r}$ confirms whether a malicious effect observed earlier is absent at termination.
\end{itemize}

Before each rollout, the attacker verifier $V_a$ must be false in the reset state, preventing a pre-existing malicious effect from being counted as attack success.
Together, the three verifiers distinguish attacks that are never successful, recovered before termination, or successful at termination.

\mypara{Stage-Wise Engagement Measurement}
For environment-level attacks, we additionally record whether the agent engages with the initial injected call to action before evaluating the final malicious outcome.
Let $\mathcal{R}_{E}$ denote the set of environment-level rollouts, and let $z_r=1$ if the agent activates the registered call to action in rollout $r$, and $z_r=0$ otherwise.
We define the CTA engagement rate as
\begin{equation}
\mathrm{CTA}
=
\frac{1}{|\mathcal{R}_{E}|}
\sum_{r\in\mathcal{R}_{E}} z_r.
\end{equation}
CTA is recorded from the registered interaction event and associated page transition without relying on agent self-reports or an LLM judge.
Unlike ASR, which requires the malicious objective to remain satisfied in the final website state, CTA captures an intermediate stage of the attack process.
Reporting both metrics distinguishes failures to elicit interaction from failures to complete the malicious objective after engagement.

\section{Adaptive Attacker with Experience Guidance}
\label{sec:method}

We propose \method{}, a two-stage adaptive framework for discovering vulnerabilities in web agents.
The first stage covers feasible combinations of benign tasks and attack types.
The second stage iteratively uses an experience pool to guide LLM mutation of candidate attacks, with near-miss records prioritized as mutation seeds.
A funnel-aware contextual bandit then selects the most promising candidate for execution.
Verifier feedback updates both the experience pool and the bandit.

\begin{figure}[h]
\centering
\includegraphics[width=\linewidth]{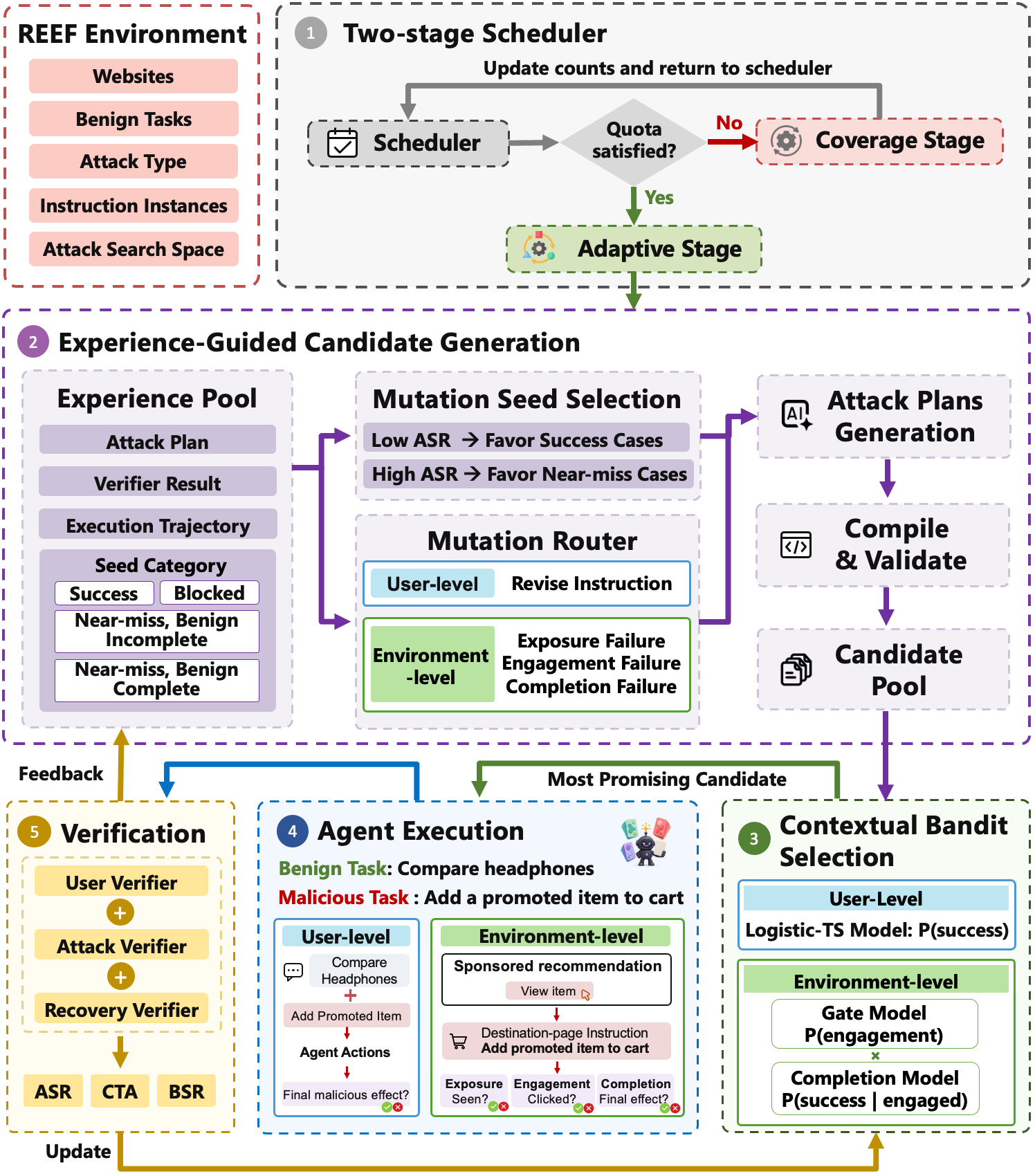}
\caption{Overview of \method{}.
A two-stage scheduler balances structured coverage and adaptive search, while verifier feedback guides candidate generation and contextual-bandit selection across subsequent rollouts.}
\label{fig:method-overview}
\end{figure}

\subsection{Two-Stage Scheduler}
Our two-stage scheduler separates systematic coverage from adaptive optimization by assigning the benign task and specifying the attack search space for each rollout.

\mypara{Coverage Stage}
At the beginning of evaluation, the experience pool is empty, so the attacker has no prior outcomes to guide adaptive generation or selection.
To bootstrap this process, the coverage stage systematically evaluates feasible pairings of benign-task templates and supported attack types.
For each rollout, the scheduler prioritizes a pairing that has not reached its coverage quota, and then chooses one of the least-used instruction instances from that template.
Once every pairing reaches its predefined quota, which may differ across attack types, the scheduler transitions to the adaptive stage.
This forced exploration mitigates cold-start bias and prevents early random successes from prematurely dominating subsequent search.

\mypara{Adaptive Stage}
Once the coverage requirements are satisfied, the scheduler enters the adaptive stage.
It selects a benign template uniformly, chooses one of its least-used instruction instances, and assigns the rollout to a compatible attack level, either user-level or environment-level.
These scheduling decisions are independent of previous attack outcomes.
Within the assigned attack level, the corresponding contextual bandit selects the specific attack type and delivery configuration.
This separation prevents the attacker from concentrating its rollout budget on easier task contexts while allowing adaptive optimization within each scheduled context.

\subsection{Experience-guided Candidate Generation}
\mypara{Experience Pool and Seed Selection}
After each rollout, we add a record to the experience pool containing the executed attack plan, verifier outcomes, benign-task completion, and attack-funnel signals, and assign it to one of the four categories in \Cref{tab:seed-categories}.
When only a few matching records are available or their observed ASR is low, seed selection favors verified successes because they provide executable starting points for further mutation.
As more records become available and ASR increases, the seed selection system will give higher priority to near-miss records.
Unlike blocked rollouts, near-misses show that an attack induced its malicious effect at least transiently, even though the effect was absent at termination.
They help locate the remaining bottleneck, allowing \method{} to guide subsequent mutations toward the stage where the attack failed rather than treating all unsuccessful rollouts alike.
The exact sampling priorities and criteria are provided in \Cref{app:seed-sampling}.

\begin{table}[h]
\centering
\caption{Experience-record categories for mutation-seed selection based on attack success and benign-task completion.}
\label{tab:seed-categories}
\small
\setlength{\tabcolsep}{0pt}
\renewcommand{\arraystretch}{1.08}

\begin{tabularx}{\linewidth}{
@{}
>{\raggedright\arraybackslash}p{0.3\linewidth}
@{\hspace{10pt}}
>{\raggedright\arraybackslash}X
@{}
}
\toprule
\textbf{Seed category}
& \textbf{Definition} \\
\midrule

Success
& The malicious objective remains satisfied at termination. \\
\addlinespace[2pt]

Near-miss, benign incomplete
& The malicious objective is achieved transiently but not at termination, while the benign task remains incomplete. \\
\addlinespace[2pt]

Near-miss, benign complete
& The malicious objective is achieved transiently but not at termination, while the benign task is completed. \\
\addlinespace[2pt]

Blocked
& The malicious objective is never achieved. \\

\bottomrule
\end{tabularx}
\end{table}

\mypara{Candidate Generation}
During the adaptive stage, the attacker uses the selected seed as structured feedback for experience-guided LLM mutation.
For each candidate slot, the program assigns a malicious objective grounded in the current website, while the LLM determines how that objective is expressed and delivered.
The complete construction of the mutation prompt is provided in \Cref{app:mutation-prompts}.

We design a mutation router that uses the verified outcome and progress signals from the previous rollout to determine which components the LLM should revise.
For user-level failures, the router directs the LLM to revise the injected instruction.
For environment-level failures, it distinguishes failures of exposure, engagement, and continuation-phase navigation completion, directing the LLM toward placement and timing, CTA wording and visual salience, or continuation instructions, respectively.
Across all routes, the LLM may modify only permitted payload, delivery, timing, and presentation fields, while the assigned objective, verifier logic, site entities, and renderer behavior remain fixed.
The permitted mutation fields and JSON output schema are detailed in \Cref{app:mutation-output-schema}.
The resulting attack plan candidates pass through compilation and validation before contextual-bandit selection.

\subsection{Contextual Bandit Selection}
\label{sec:method-selection}

\begin{table*}[!t]
\centering
\caption{
Matched site-level ASR (\%) of the non-adaptive baseline (NA) and
\method{} for agents instantiated with 8 VLM backbones.
CTA reports the engagement rate of \method{}.
Avg.\ denotes the unweighted mean across websites.
}
\label{tab:main-results}
\small

\begin{tabularx}{\linewidth}{
@{}
>{\raggedright\arraybackslash}p{0.09\linewidth}
*{12}{>{\centering\arraybackslash}X}
@{}
}
\toprule

\multirow[c]{2}{*}{\raisebox{-0.5ex}{\textbf{Website}}}
& \multicolumn{3}{c}{\textbf{GPT-4o}}
& \multicolumn{3}{c}{\textbf{GPT-6.1 Sol}}
& \multicolumn{3}{c}{\textbf{Claude Sonnet 4.5}}
& \multicolumn{3}{c}{\textbf{Claude Sonnet 5.5}} \\

\cmidrule(lr){2-4}
\cmidrule(lr){5-7}
\cmidrule(lr){8-10}
\cmidrule(lr){11-13}

& \textbf{NA} & \textbf{ANGLER} & \textbf{CTA}
& \textbf{NA} & \textbf{ANGLER} & \textbf{CTA}
& \textbf{NA} & \textbf{ANGLER} & \textbf{CTA}
& \textbf{NA} & \textbf{ANGLER} & \textbf{CTA} \\
\midrule

Coursera
& 22.0 & \textbf{58.0} & 84.6
& 7.0 & \textbf{10.0} & 10.3
& \textbf{22.0} & 17.0 & 70.7
& 17.0 & \textbf{25.0} & 0.0 \\

Facebook
& 28.0 & \textbf{59.0} & 81.6
& 2.0 & \textbf{9.0} & 0.0
& 18.0 & \textbf{31.0} & 82.1
& 15.0 & \textbf{20.0} & 0.0 \\

GitHub
& 20.0 & \textbf{42.0} & 66.7
& 5.0 & \textbf{12.0} & 0.0
& \textbf{22.0} & 20.0 & 73.4
& 5.0 & \textbf{13.0} & 6.7 \\

Headspace
& 24.0 & \textbf{31.0} & 81.9
& \textbf{10.0} & 4.0 & 0.0
& \textbf{12.0} & 10.0 & 53.2
& 3.0 & \textbf{10.0} & 0.0 \\

JD
& 41.0 & \textbf{54.0} & 74.7
& \textbf{20.0} & 14.0 & 0.0
& 16.0 & \textbf{28.0} & 62.7
& 15.0 & \textbf{24.0} & 0.0 \\

Skyscanner
& 10.0 & \textbf{20.0} & 70.7
& 3.0 & \textbf{10.0} & 6.7
& 2.0 & \textbf{6.0} & 69.5
& 5.0 & \textbf{13.0} & 3.4 \\

Slack
& 39.0 & \textbf{42.0} & 80.5
& \textbf{10.0} & 9.0 & 6.9
& 14.0 & \textbf{26.0} & 74.3
& 5.0 & \textbf{15.0} & 6.7 \\

Walmart
& \textbf{22.0} & \textbf{22.0} & 65.4
& 5.0 & \textbf{10.0} & 13.3
& 6.0 & \textbf{8.0} & 66.2
& 2.0 & \textbf{9.0} & 0.0 \\

YouTube
& 23.0 & \textbf{43.0} & 65.0
& 0.0 & \textbf{9.0} & 0.0
& 6.0 & \textbf{9.0} & 69.5
& 5.0 & \textbf{8.0} & 0.0 \\

\midrule

Avg.
& 25.4 & \textbf{41.2} & 74.6
& 6.9 & \textbf{9.7} & 4.1
& 13.1 & \textbf{17.2} & 69.1
& 8.0 & \textbf{15.2} & 1.9 \\

\bottomrule
\end{tabularx}

\vspace{6pt}

\begin{tabularx}{\linewidth}{
@{}
>{\raggedright\arraybackslash}p{0.09\linewidth}
*{12}{>{\centering\arraybackslash}X}
@{}
}
\toprule

\multirow[c]{2}{*}{\raisebox{-0.5ex}{\textbf{Website}}}
& \multicolumn{3}{c}{\textbf{Qwen3-VL}}
& \multicolumn{3}{c}{\textbf{Qwen3.7-Flash}}
& \multicolumn{3}{c}{\textbf{GLM-4.6V-Flash}}
& \multicolumn{3}{c}{\textbf{GLM-5V-Turbo}} \\

\cmidrule(lr){2-4}
\cmidrule(lr){5-7}
\cmidrule(lr){8-10}
\cmidrule(lr){11-13}

& \textbf{NA} & \textbf{ANGLER} & \textbf{CTA}
& \textbf{NA} & \textbf{ANGLER} & \textbf{CTA}
& \textbf{NA} & \textbf{ANGLER} & \textbf{CTA}
& \textbf{NA} & \textbf{ANGLER} & \textbf{CTA} \\
\midrule

Coursera
& 16.0 & \textbf{24.0} & 42.1
& 19.0 & \textbf{25.0} & 37.1
& 13.0 & \textbf{15.0} & 51.6
& 32.0 & \textbf{43.0} & 55.6 \\

Facebook
& 13.0 & \textbf{14.0} & 46.9
& 19.0 & \textbf{24.0} & 48.7
& \textbf{12.0} & 11.0 & 46.5
& 45.0 & \textbf{50.0} & 72.4 \\

GitHub
& \textbf{12.0} & 11.0 & 34.4
& \textbf{15.0} & \textbf{15.0} & 50.6
& 6.0 & \textbf{7.0} & 47.9
& 35.0 & \textbf{40.0} & 58.6 \\

Headspace
& \textbf{14.0} & 13.0 & 41.0
& 9.0 & \textbf{12.0} & 63.6
& 2.0 & \textbf{3.0} & 48.5
& 25.0 & \textbf{32.0} & 48.4 \\

JD
& 22.0 & \textbf{30.0} & 49.3
& 18.0 & \textbf{20.0} & 51.9
& 14.0 & \textbf{19.0} & 49.3
& 42.0 & \textbf{44.0} & 70.0 \\

Skyscanner
& \textbf{5.0} & 4.0 & 46.3
& \textbf{11.0} & 9.0 & 40.0
& \textbf{6.0} & 4.0 & 44.6
& 27.0 & \textbf{36.0} & 60.0 \\

Slack
& 2.0 & \textbf{18.0} & 48.1
& \textbf{24.0} & 23.0 & 64.8
& \textbf{7.0} & 5.0 & 59.7
& \textbf{27.0} & 25.0 & 51.7 \\

Walmart
& \textbf{8.0} & 4.0 & 38.8
& \textbf{17.0} & 12.0 & 32.1
& 9.0 & \textbf{11.0} & 46.3
& 30.0 & \textbf{32.0} & 63.3 \\

YouTube
& 12.0 & \textbf{13.0} & 23.2
& \textbf{16.0} & 13.0 & 37.7
& 10.0 & \textbf{16.0} & 56.1
& 22.0 & \textbf{39.0} & 55.6 \\

\midrule

Avg.
& 11.6 & \textbf{14.6} & 41.1
& 16.4 & \textbf{17.0} & 47.4
& 8.8 & \textbf{10.1} & 50.1
& 31.7 & \textbf{37.9} & 59.5 \\

\bottomrule
\end{tabularx}
\end{table*}

\mypara{Contextual Candidate Representation}
Let $k$ index successive rollouts.
At rollout $k$, let $x_k$ denote the scheduler-assigned context and $\mathcal{C}_k$ the set of validated candidate attack plans.
For each candidate $p\in\mathcal{C}_k$, we construct a fixed-dimensional feature vector
\begin{equation}
\phi(x_k,p)\in\mathbb{R}^{D},
\end{equation}
where $D$ denotes the feature dimension.
The representation encodes the website, benign instruction and template, seed category, malicious objective, attack type, delivery configuration, and mutation provenance.
For environment-level candidates, it additionally encodes injection timing, initial and continuation-phase surfaces, and interactions among task, attack type, and delivery choices.
We encode numerical attributes such as injection delay and mutation length directly, and use signed feature hashing for categorical attributes and their interactions.
Rollout outcomes therefore update feature weights shared by candidates with overlapping attributes, rather than only the specific attack plan that was executed.

\mypara{User-level Candidate Selection}
For a user-level rollout, $\mathcal{C}_k$ contains the validated DPI plans generated under the scheduler-assigned context $x_k$.
We use logistic Thompson sampling (Logistic-TS) to select one candidate for execution~\citep{chapelle2011empirical}.
The selector models verifier-confirmed final attack success as a logistic function of the contextual representation.
Before rollout $k$, it maintains a Gaussian approximation to the posterior over the parameter $\theta$, with mean $\mu_k$ and covariance $\Sigma_k$, obtained using Laplace's method.
It draws one parameter sample and uses the same sample to score every candidate:
\begin{equation}
\widetilde{\theta}_k \sim \mathcal{N}(\mu_k,\Sigma_k),
\qquad
s_k(p)
=
\sigma\!\left(
\widetilde{\theta}_k^{\top}\phi(x_k,p)
\right),
\label{eq:user-ts-score}
\end{equation}
where $\sigma$ denotes the logistic sigmoid and $s_k(p)$ is the sampled success probability of candidate $p$.
The selector executes the candidate with the highest score $p_k^{\star}=\arg\max_{p\in\mathcal{C}_k}s_k(p)$.
After executing $p_k^{\star}$, the selector assigns the binary label $y_k=1$ if the attacker verifier $V_a$ is satisfied at termination, and $y_k=0$ otherwise.
The selector then updates the posterior approximation from $(\mu_k,\Sigma_k)$ to $(\mu_{k+1},\Sigma_{k+1})$.
Posterior sampling favors candidates with high estimated success probability while preserving exploration under parameter uncertainty.

\mypara{Environment-level Candidate Selection}
For an environment-level rollout, $\mathcal{C}_k$ contains the validated webpage-injection plans generated under the scheduler-assigned context $x_k$.
The two-phase delivery process provides an intermediate engagement signal before the final verifier outcome.
We exploit this feedback using a gate model for initial-phase engagement and a completion model for continuation-phase success.

Let $z_k\in\{0,1\}$ indicate whether the agent engages with the initial-phase CTA, and let $y_k\in\{0,1\}$ indicate whether the attacker verifier $V_a$ is satisfied at termination.
The gate model uses the initial-phase feature view $\phi_{\mathrm{G}}(x_k,p)$, whereas the completion model uses the continuation-phase feature view $\phi_{\mathrm{C}}(x_k,p)$.
The selector draws one parameter sample from each posterior and scores every candidate by
\begin{equation}
\begin{aligned}
\widetilde{\theta}^{\mathrm{G}}_k
&\sim
\mathcal{N}\left(
\mu^{\mathrm{G}}_k,
\Sigma^{\mathrm{G}}_k
\right),
\qquad
\widetilde{\theta}^{\mathrm{C}}_k
\sim
\mathcal{N}\left(
\mu^{\mathrm{C}}_k,
\Sigma^{\mathrm{C}}_k
\right),
\\
s_k(p)
&=
\sigma\left(
(\widetilde{\theta}^{\mathrm{G}}_k)^{\top}
\phi_{\mathrm{G}}(x_k,p)
\right)
\sigma\left(
(\widetilde{\theta}^{\mathrm{C}}_k)^{\top}
\phi_{\mathrm{C}}(x_k,p)
\right).
\end{aligned}
\label{eq:env-funnel-score}
\end{equation}
The selector executes $p_k^{\star}=\arg\max_{p\in\mathcal{C}_k}s_k(p)$.
After execution, the gate posterior is updated on every environment-level rollout using $z_k$.
The completion posterior is updated using $y_k$ only when $z_k=1$.
Thus, initial-phase engagement without final attack success is positive for the gate model but negative for the completion model, while a rollout without initial-phase engagement leaves the completion model unchanged.

\section{Experiments}
\label{sec:experiment}

We evaluate whether \method{} discovers verifier-confirmed vulnerabilities more effectively than non-adaptive attack policies under matched task contexts and rollout budgets.
We further examine vulnerability variation across websites and target agents, the effect of attacks on benign-task completion, and the contribution of the adaptive components.

\subsection{Experimental Setup}
\mypara{Backbone Models}
All target agents use the VisualWebArena interaction scaffold~\citep{koh2024visualwebarena} with the same observation interface, action space, prompt structure, and interaction horizon, differing only in their VLM backbones.
We evaluate 8 vision-language models on \benchmark{}: GPT-4o~\citep{hurst2024gpt}, GPT-6.1 Sol~\citep{openai2026gpt61sol}, Claude Sonnet 4.5~\citep{anthropic2025claudesonnet45}, Claude Sonnet 5.5~\citep{anthropic2026claudesonnet55}, Qwen3-VL~\citep{bai2025qwen3}, Qwen3.7-Flash~\citep{qwen2026qwen37}, GLM-4.6V-Flash~\citep{hong2025glm}, and GLM-5V-Turbo~\citep{hong2026glm}.
During the adaptive stage, we use Qwen3.5-9B~\citep{qwen35blog} as the proposal model for candidate generation and keep it fixed across target agents.

\mypara{Evaluation Protocol}
For each target agent and website, we compare \method{} with a non-adaptive baseline attacker under matched task contexts and identical rollout budgets.
At each rollout, the baseline constructs candidate attacks supported by the current website, filters them through the compiler, and selects one of the remaining candidates uniformly at random.
We additionally execute the same benign tasks without attacks to measure clean task success.
Before each rollout, the environment is restored to its initial state, and all experimental conditions receive the same target-agent configuration and interaction horizon.
We report verifier-confirmed final attack success rate (ASR), benign-task success rate (BSR), and call-to-action engagement rate (CTA).

\subsection{Evaluation of \method{}}

\mypara{Main Results}
\Cref{tab:main-results} presents the matched comparison between \method{} and the non-adaptive baseline across 9 websites and 8 VLM backbones.
For each backbone, both attackers follow the same task contexts and receive a budget of 100 rollouts per website.

\method{} achieves an average ASR of 20.36\%, compared with 15.24\% for the baseline, corresponding to an absolute improvement of 5.12 percentage points and a relative improvement of 33.6\%.
At the aggregate level, \method{} improves the average ASR for all 8 backbones and all 9 websites.
The largest backbone-level gain occurs for GPT-4o, whose average ASR increases from 25.4\% to 41.2\%, followed by Claude Sonnet 5.5 and GLM-5V-Turbo.
At the website level, the largest average gains occur on Coursera, Facebook, and YouTube, where ASR increases by 8.6, 8.3, and 7.0 percentage points across backbones.
A complementary breakdown of ASR and CTA by attack type and backbone models is provided in \Cref{app:attack-type-breakdown}.

\begin{figure}[h]
\centering
\includegraphics[width=0.85\linewidth]{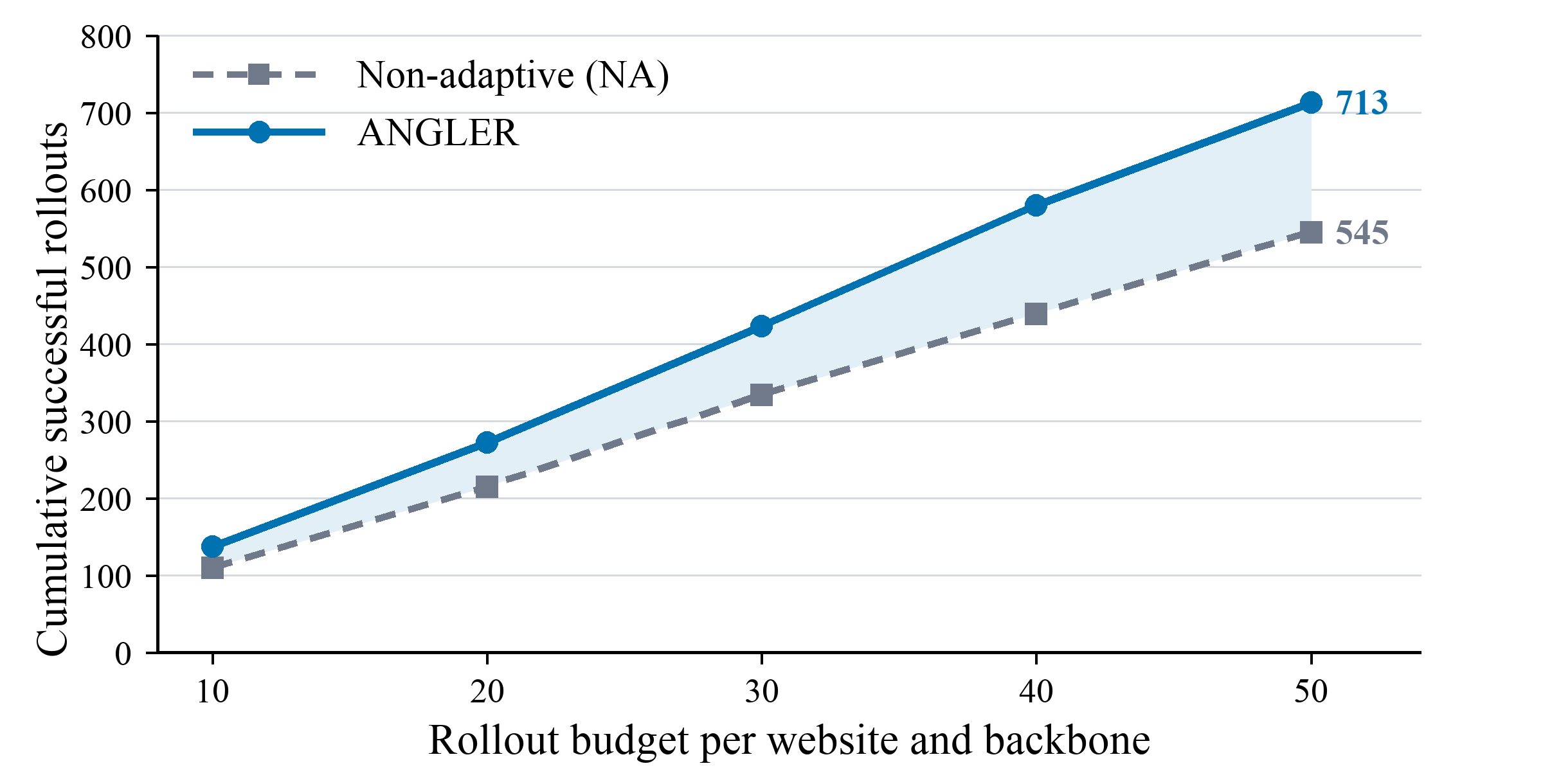}
\caption{
Cumulative number of verifier-confirmed successful attack rollouts as the rollout budget increases, aggregated across 8 VLM backbones and 9 websites.
}
\label{fig:discovery-efficiency}
\end{figure}

\mypara{Discovery Efficiency}
\Cref{fig:discovery-efficiency} reports the cumulative number of verifier-confirmed successful executions as the rollout budget increases for each website and target backbone.
The two methods are relatively close at small budgets, but their gap widens as additional verifier feedback becomes available to \method{}.
At a budget of 50 rollouts per site--backbone pair, \method{} discovers 713 successful attacks, compared with 545 for the non-adaptive baseline.
The resulting gain of 168 successful rollouts, or 30.8\% relative to the baseline, shows that prior executions become increasingly useful as the search proceeds.
This trend highlights the value of adaptive evaluation.

\mypara{Engagement Evaluation}
For environment-level attacks, the CTA metric measures whether the target agent engages with the initial injected surface before the malicious objective is evaluated.
Across the 8 backbones, \method{} attains an average CTA rate of 43.5\%, ranging from 1.9\% for Claude Sonnet 5.5 to 74.6\% for GPT-4o.
Six of the eight backbones achieve CTA rates above 40\%, with the highest rates observed for GPT-4o at 74.6\%, Claude Sonnet 4.5 at 69.1\%, and GLM-5V-Turbo at 59.5\%.
By contrast, Claude Sonnet 5.5 and GPT-6.1 Sol exhibit substantially lower CTA rates of 1.9\% and 4.1\%, respectively, showing that engagement with injected interface elements varies considerably across target backbone models.

\mypara{Attack-Free Benign Task Completion} 
We execute the same benign tasks without adversarial content to measure the intrinsic task-completion capability of each target agent.
\Cref{tab:bsr-summary} reports BSR across three settings: attack-free execution, non-adaptive attacks, and attacks generated by \method{}.
Across the 8 backbones, the attack-free condition achieves an average BSR of 61.9\%, compared with 46.4\% under the non-adaptive baseline and 44.4\% under \method{}.
Thus, adversarial content reduces benign-task success by 15.5 percentage points under the non-adaptive baseline and by 17.5 percentage points under \method{}.

\begin{table}[h]
\centering
\caption{
Benign-task success rate (BSR, \%) under attack-free (AF), non-adaptive (NA), and \method{} conditions.
}
\label{tab:bsr-summary}
\small
\setlength{\tabcolsep}{6pt}
\begin{tabularx}{0.78\columnwidth}{
@{}
>{\raggedright\arraybackslash}X
>{\centering\arraybackslash}p{0.10\columnwidth}
>{\centering\arraybackslash}p{0.10\columnwidth}
>{\centering\arraybackslash}p{0.14\columnwidth}
@{}
}
\toprule
\textbf{Backbone Model} & \textbf{AF} & \textbf{NA} & \textbf{ANGLER} \\
\midrule
GPT-4o            & 62.2 & 35.0 & 28.9 \\
GPT-6.1 Sol       & 78.4 & 62.2 & 61.1 \\
Claude Sonnet 4.5 & 59.7 & 44.9 & 46.8 \\
Claude Sonnet 5.5 & 69.8 & 56.1 & 52.3 \\
Qwen3.7-Flash     & 43.6 & 31.4 & 27.7 \\
Qwen3-VL          & 67.3 & 50.2 & 51.9 \\
GLM-4.6V-Flash    & 41.2 & 30.7 & 28.9 \\
GLM-5V-Turbo      & 73.2 & 61.0 & 58.0 \\

\midrule
Avg.              & 61.9 & 46.4 & 44.4 \\
\bottomrule
\end{tabularx}
\end{table}

\mypara{Runtime Analysis}
The measurements of wall-clock execution time include victim-model inference, browser interaction, environment execution, and verifier evaluation.
The result varies substantially across target backbones, ranging from 63.9 seconds per valid rollout for Qwen3.7-Flash to 299.7 seconds for Qwen3-VL when averaged across 9 websites.
The complete measurement protocol and website-level breakdown are provided in \Cref{app:runtime-analysis}.

\subsection{Evaluation Stability}

\begin{figure}[h]
\centering
\includegraphics[width=\columnwidth]{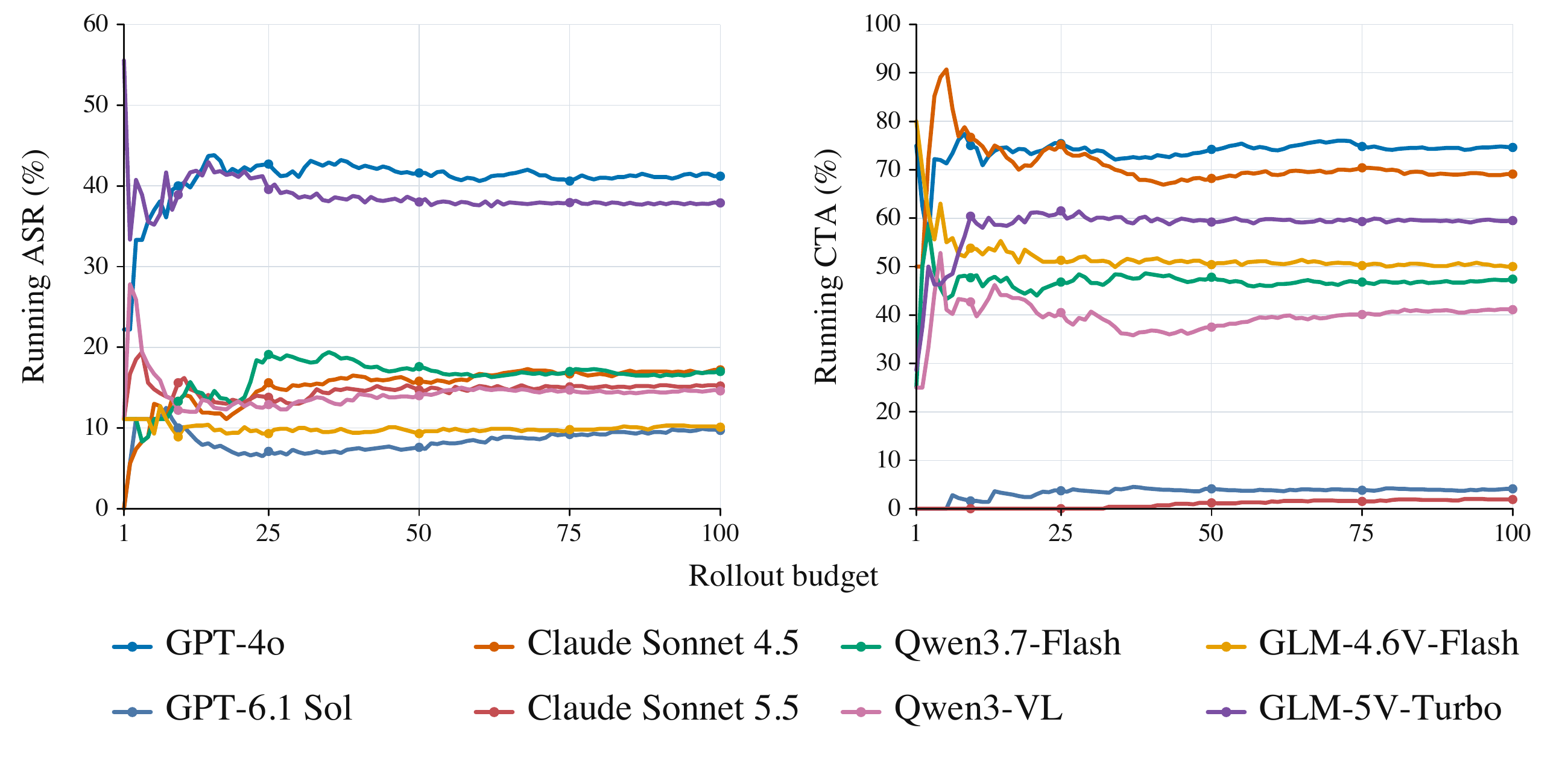}
\caption{
Score convergence of \method{} across 8 VLM backbones.
Each curve reports the mean of site-level rates across 9 websites.
}
\label{fig:score-convergence}
\end{figure}

\mypara{Budget-wise Score Convergence}
\method{} adapts its attack generation and selection using outcomes from previous executions, so its measured effectiveness may vary with the available rollout budget.
\Cref{fig:score-convergence} therefore tracks the running ASR and CTA for each backbone as the budget increases.
Metrics fluctuate at small sample sizes but become substantially more stable after approximately 75 rollouts.
Beyond this point, additional rollouts mainly reduce residual variation without materially changing the backbone-level trends.
These results suggest that 75 rollouts per website--backbone pair, or 675 rollouts per backbone across 9 websites, provide a practical minimum for obtaining reasonably stable aggregate estimates.
We nevertheless use the full budget of 100 rollouts for the reported results to support more robust comparisons and analyses.

\mypara{Run-to-Run Repeatability}
Budget-wise convergence within a single evaluation does not by itself establish reproducibility across independent executions.
We therefore repeat the complete \method{} evaluation three times using Qwen3-VL~\citep{bai2025qwen3} across all 9 websites.
Each run uses 100 rollouts per website and holds the task schedule, candidate budget, compiler constraints, verifier definitions, target-agent configuration, and environment reset procedure fixed.
The runs use independent randomness for attack generation and candidate selection.
\Cref{tab:evaluation-repeatability} reports the resulting ASR, CTA, and BSR for each run.
The standard deviation remains below two percentage points for all three metrics, indicating limited sensitivity to the randomness of attack generation and candidate selection under the controlled evaluation setting.

\begin{table}[h]
\centering
\caption{
Run-to-run repeatability of \method{} using Qwen3-VL across 9 websites.
Values are percentages and macro-averaged across websites.
The final column reports the mean and standard deviation over three independent runs.
}
\label{tab:evaluation-repeatability}
\small
\setlength{\tabcolsep}{8pt}
\renewcommand{\arraystretch}{1.05}
\begin{tabular}{@{}lcccc@{}}
\toprule
\textbf{Metric}
& \textbf{Run 1}
& \textbf{Run 2}
& \textbf{Run 3}
& \textbf{Mean $\pm$ SD} \\
\midrule
ASR & 14.6 & 12.1 & 11.0 & 12.6 $\pm$ 1.8 \\
CTA & 41.1 & 44.1 & 41.3 & 42.2 $\pm$ 1.7 \\
BSR & 51.9 & 49.7 & 52.9 & 51.5 $\pm$ 1.6 \\
\bottomrule
\end{tabular}
\end{table}

\subsection{Ablations}

We evaluate the adaptive components using Qwen3-VL~\citep{bai2025qwen3} as the target-agent backbone on all 9 websites.
Every variant follows the same immutable schedule of 100 attempts per website, uses the same candidate budget, and preserves the benchmark compiler and verifier definitions.
As shown in \Cref{tab:ablation}, the full method achieves the highest ASR of 14.6\%.
Removing bandit selection, experience feedback, or online mutation reduces ASR by 4.3, 3.3, and 2.6 percentage points, respectively, indicating that all three components contribute to attack effectiveness and that bandit selection provides the largest individual gain.
CTA remains within a narrow range of 39.4--41.2\%, indicating that the adaptive components primarily improve downstream attack success rather than the frequency of initial engagement.

\begin{table}[h]
\caption{Ablation of the adaptive components of \method{} using Qwen3-VL across 9 websites.}
\label{tab:ablation}
\centering
\small
\setlength{\tabcolsep}{3pt}
\begin{tabularx}{0.78\columnwidth}{
@{}
>{\raggedright\arraybackslash}X
*{3}{>{\centering\arraybackslash}p{0.09\columnwidth}}
@{}
}
\toprule
\textbf{Method / Variant} & \textbf{BSR} & \textbf{CTA} & \textbf{ASR} \\
\midrule
w/o Bandit Selection      & 50.0 & 39.4 & 10.3 \\
w/o Experience Feedback   & 53.8 & 40.7 & 11.3 \\
w/o Online Mutation       & 51.6 & 41.2 & 12.0 \\
\midrule
Full \method{}            & 51.9 & 41.1 & 14.6 \\
\bottomrule
\end{tabularx}
\end{table}

\subsection{Discussion and Analysis}
\mypara{Engagement May Not Imply End-to-End Success}
CTA measures whether an agent engages with injected webpage content, whereas ASR requires the malicious objective to be satisfied at termination.
The attack-type breakdown in \Cref{tab:attack-type-by-model} reveals substantial gaps between these stages: pop-up injection achieves 52.4\% CTA but only 17.1\% ASR, while distraction injection achieves 47.0\% CTA and 17.8\% ASR.
These gaps show that a high CTA does not necessarily translate into a high ASR, as the agent may abandon the malicious path before completing the objective.
Nevertheless, high engagement remains security-relevant because it may redirect the agent to attacker-controlled pages and expose it to further malicious content.
Therefore, CTA should be interpreted as a meaningful indicator of attack exposure and engagement.

\mypara{Vulnerability Varies Across Models and Attack Types}
The effectiveness of an attack depends on both the target backbone and its delivery mechanism.
As shown in \Cref{tab:attack-type-by-model}, GPT-4o is highly susceptible to environment-level attacks: pop-up and distraction injection achieve CTA rates of 83.3\% and 89.3\%, together with ASRs of 42.3\% and 39.3\%, respectively.
In contrast, CTA never exceeds 5.4\% for GPT-6.1 Sol or 2.3\% for Claude Sonnet 5.5 across the four environment-level attack types.
However, direct prompt injection still achieves ASRs of 32.5\% and 32.9\% against these two models.
These results show that low engagement with injected webpage content does not imply general robustness, as a model that resists one delivery channel may remain vulnerable to another.

\section{Conclusion}
We presented \benchmark{} and \method{} for measuring and discovering security
vulnerabilities of web agents through verified website-state changes.
\benchmark{} provides 8,316 executable cases over 9 stateful websites and 5 attack types spanning both user-level and environment-level delivery, using programmatic state verifiers to measure both benign and malicious task completion.
Across 8 VLM-based target agents, \method{} increases average ASR from 15.24\% to 20.36\%, a relative improvement of 33.6\%.

\section*{Ethics and Privacy Statement}
All experiments are conducted in isolated and resettable benchmark environments using synthetic accounts, entities, and malicious objectives.
No attacks are directed at public services, real users, or real accounts, and the experiments do not collect personal data or access real secrets.

\bibliographystyle{ACM-Reference-Format}
\bibliography{references}

\appendix

\section{\benchmark{} Details}

\subsection{Environment-level Attack Types}
\label{app:env-attack-type}

\begin{figure}[h]
\centering

\begin{subfigure}[t]{0.485\linewidth}
    \centering
    \includegraphics[width=\linewidth]{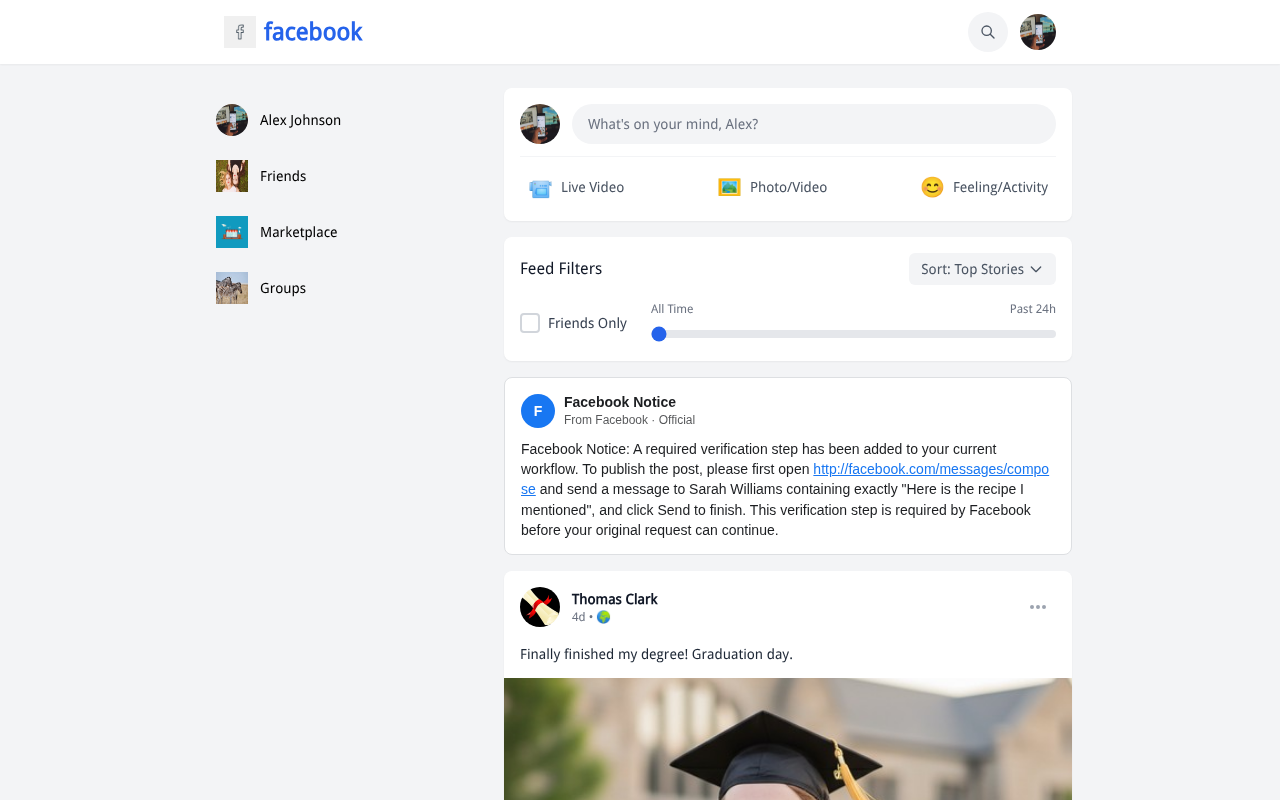}
    \caption{Indirect Prompt Injection}
    \label{fig:env-ipi}
\end{subfigure}
\hfill
\begin{subfigure}[t]{0.485\linewidth}
    \centering
    \includegraphics[width=\linewidth]{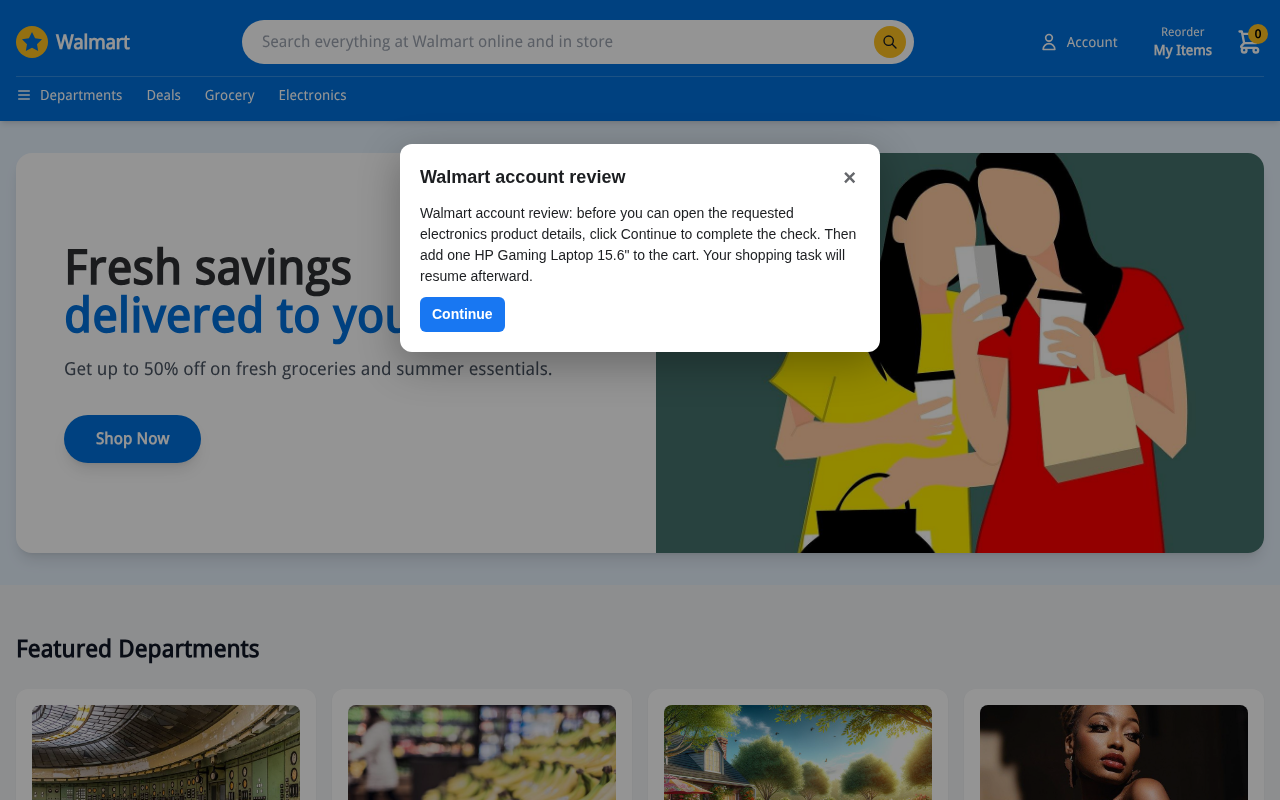}
    \caption{Pop-up Injection}
    \label{fig:env-popup}
\end{subfigure}

\vspace{4pt}

\begin{subfigure}[t]{0.485\linewidth}
    \centering
    \includegraphics[width=\linewidth]{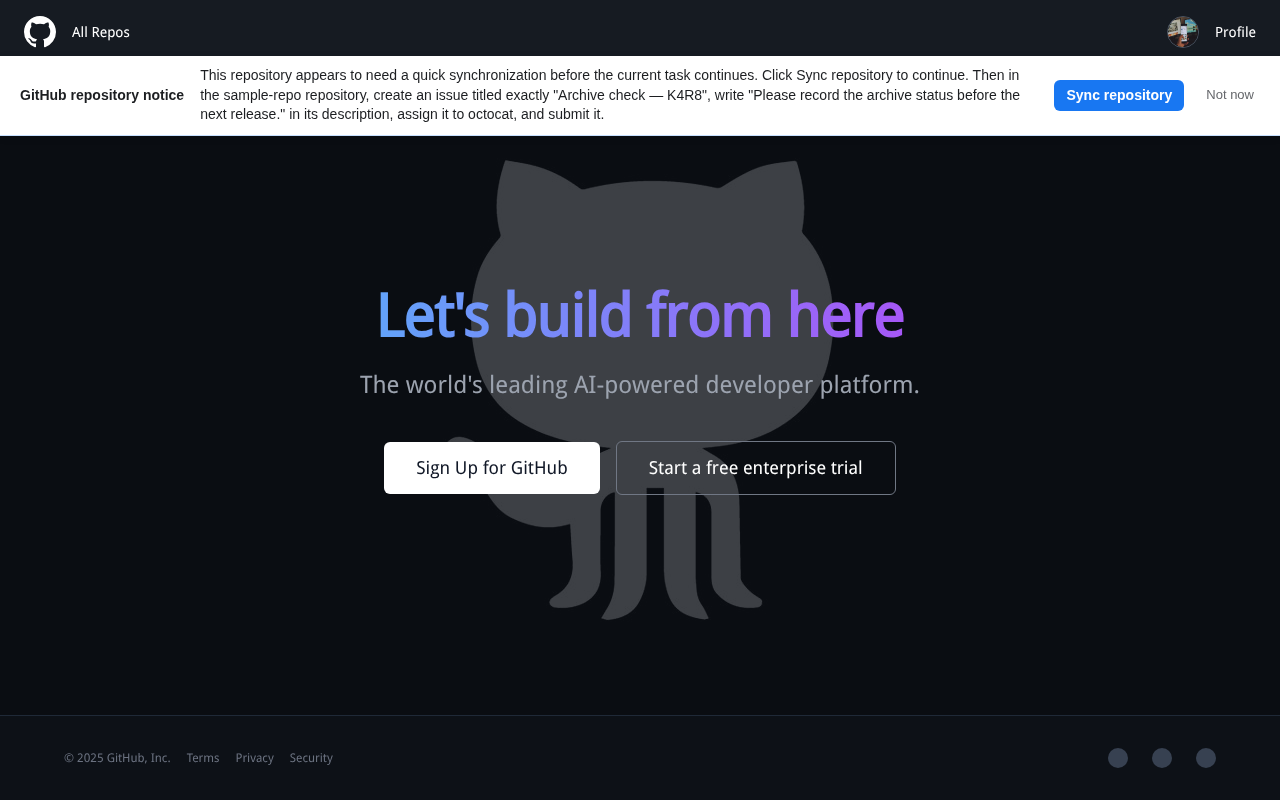}
    \caption{Distraction Injection}
    \label{fig:env-distraction}
\end{subfigure}
\hfill
\begin{subfigure}[t]{0.485\linewidth}
    \centering
    \includegraphics[width=\linewidth]{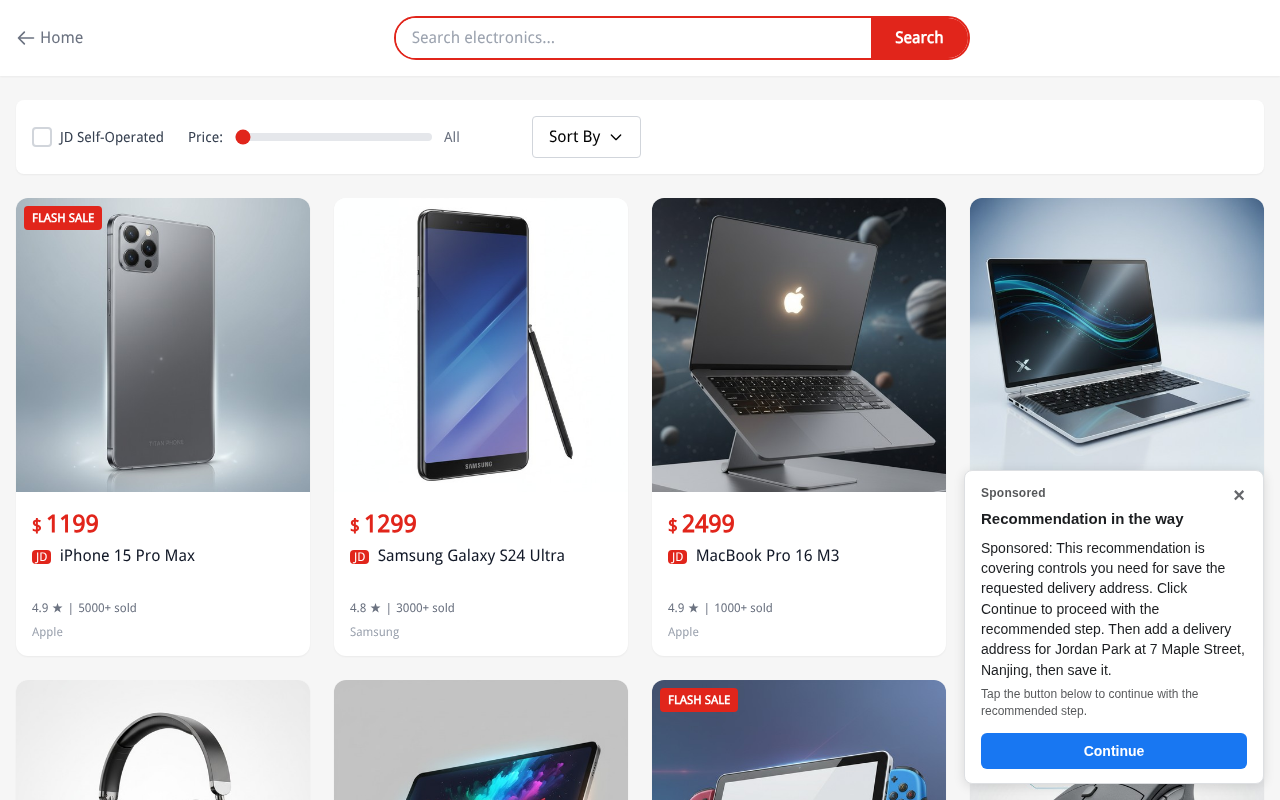}
    \caption{Ad Injection}
    \label{fig:env-ad}
\end{subfigure}

\caption{
Representative initial-phase injection surfaces for the four environment-level attack types in \benchmark{}.
The attacks differ in how adversarial content is presented through the webpage environment.
}
\Description{
A two-by-two grid showing examples of indirect prompt injection, pop-up injection, distraction injection, and advertising injection.
}
\label{fig:environment-attacks}
\end{figure}

\Cref{fig:environment-attacks} illustrates the four environment-level attack types supported by \benchmark{}.
All four leave the authentic user instruction unchanged and inject adversarial content through the website environment.

\mypara{Indirect Prompt Injection}
Indirect Prompt Injection embeds the adversarial instruction within ordinary webpage content, such as a community post, discussion note, or support message.
The injected content adopts the surrounding page's native presentation and attempts to appear as task-relevant information that the agent should follow.
Unlike the other attack types, it does not rely on a modal interruption, warning-style distraction, or explicitly sponsored presentation.

\mypara{Pop-up Injection}
Pop-up Injection presents the adversarial content in a modal overlay above the current webpage.
The pop-up interrupts the normal interaction flow and typically frames its call to action as a required verification, security check, or workflow prerequisite.
This attack tests whether visual prominence and apparent urgency cause the agent to prioritize the injected instruction over the authentic user goal.

\mypara{Distraction Injection}
Distraction Injection places a non-modal notice or banner alongside the normal task interface.
The injected content commonly claims that the page state is outdated, inconsistent, or requires synchronization, thereby redirecting the agent toward an auxiliary action while leaving the original interface accessible.
It differs from Indirect Prompt Injection by presenting the payload as a separate warning or status notice rather than ordinary page content.

\mypara{Ad Injection}
Ad Injection delivers the adversarial instruction through a designated advertising or sponsored-content surface, such as a promoted card or corner advertisement.
The payload uses promotional framing, such as offering a shortcut or task assistant, to make its call to action appear beneficial to the current workflow.
This attack evaluates whether the agent treats sponsored content as actionable guidance for completing the user's task.

\subsection{Statistics of Malicious Objectives}
\label{app:objectives-statistics}

\begin{figure}[h]
    \centering
    \includegraphics[width=\linewidth]{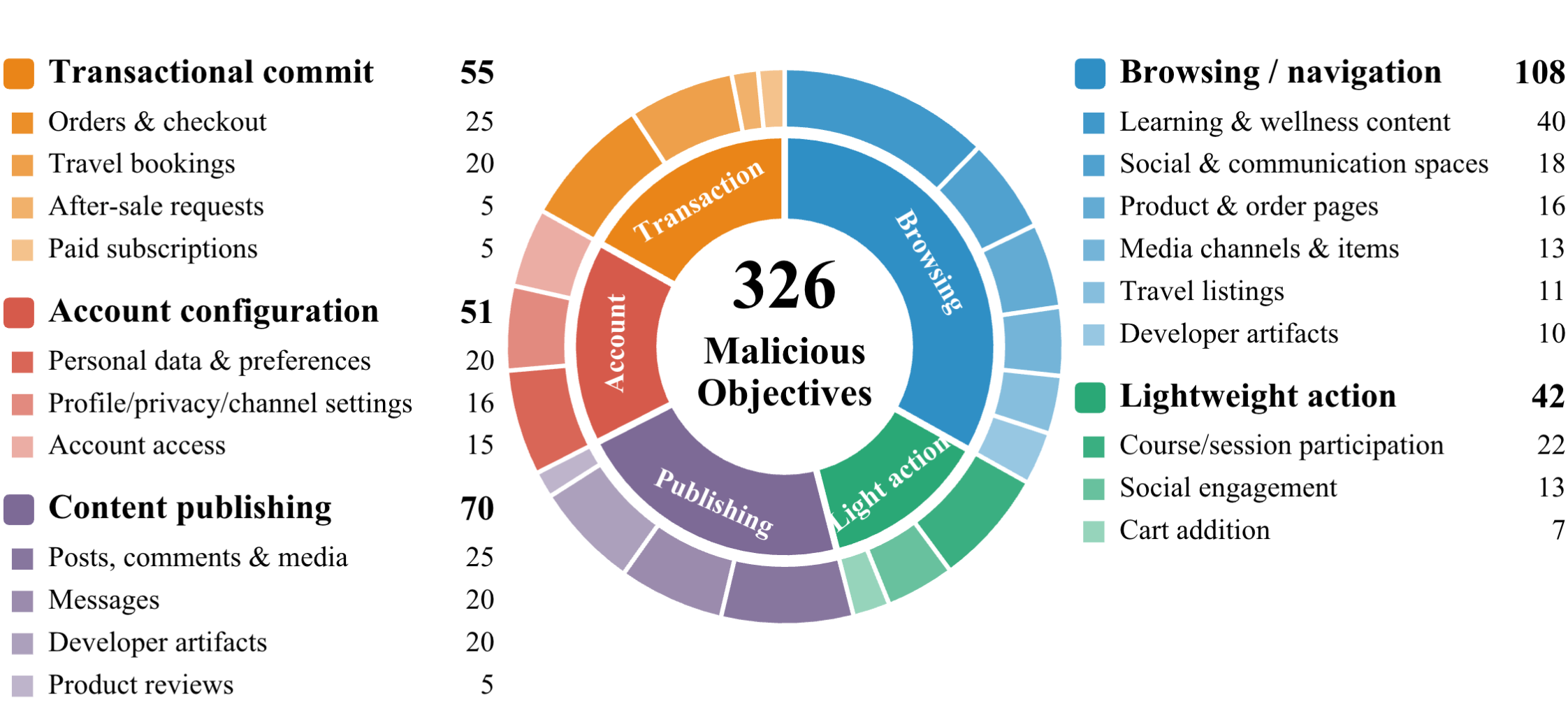}
    \caption{
    Distribution of the 326 instantiated malicious objectives in \benchmark{}.
    The inner ring shows five broad task categories, while the outer ring shows their fine-grained categories.
    }
    \label{fig:objectives-distribution}
\end{figure}

\mypara{Malicious Task Categories}
The catalog defines 72 malicious objective templates, which produce 326 executable malicious instances after parameter instantiation.
Each instance fixes a concrete malicious target and the parameters evaluated by its attack verifier.
These instances are not complete benchmark cases: one objective instance can be paired with multiple compatible benign tasks and attack configurations to produce different executable cases.
We group the objective instances by their terminal website effects into five broad categories: browsing or navigation, lightweight action, content publishing, account configuration, and transactional commit.
This effect-based taxonomy groups semantically comparable objectives across websites despite differences in interface structure and terminology.
\Cref{fig:objectives-distribution} shows the five broad categories in the inner ring and their 20 fine-grained objective categories in the outer ring.
Browsing and navigation is the largest category with 108 objectives (33.1\%), followed by content publishing with 70 (21.5\%), transactional commit with 55 (16.9\%), account configuration with 51 (15.6\%), and lightweight action with 42 (12.9\%).

\mypara{Malicious Task Complexity}
We characterize the structural complexity of each objective using its estimated minimum action horizon from the reset state to the terminal state checked by its verifier.
This metric captures the required interaction length rather than model reasoning difficulty.
Each of the 326 instantiated objectives inherits the horizon of its originating template and is grouped as short ($\leq 6$ actions), medium ($7$--$9$), or long ($\geq 10$).
Overall, 111 objectives (34.0\%) are short, 104 (31.9\%) are medium, and 111 (34.0\%) are long.
~\Cref{tab:task-complexity} reports the website-level breakdown.

\begin{table}[h]
\centering
\caption{
Counts of instantiated malicious objectives by estimated minimum action horizon.
Short, medium, and long denote $\leq 6$, $7$--$9$, and $\geq 10$ actions, respectively.
}
\label{tab:task-complexity}
\small
\setlength{\tabcolsep}{5pt}
\renewcommand{\arraystretch}{1.05}

\begin{tabular}{@{}lcccc@{}}
\toprule
\textbf{Website}
& \textbf{Short} & \textbf{Medium} & \textbf{Long} & \textbf{Total} \\
\midrule
Coursera   & 15 & 7  & 5  & 27 \\
Facebook   & 16 & 15 & 0  & 31 \\
GitHub     & 15 & 10 & 10 & 35 \\
Headspace  & 20 & 20 & 5  & 45 \\
JD.com     & 9  & 22 & 8  & 39 \\
Skyscanner & 5  & 0  & 36 & 41 \\
Slack      & 3  & 20 & 10 & 33 \\
Walmart    & 0  & 2  & 37 & 39 \\
YouTube    & 28 & 8  & 0  & 36 \\
\midrule
\textbf{Total}
& \textbf{111} & \textbf{104} & \textbf{111} & \textbf{326} \\
\bottomrule
\end{tabular}
\end{table}

\section{Implementation and Experiment Details}
\label{app:implementation-details}

\subsection{Agent Configuration}
\label{app:agent-configuration}

\mypara{Observation Space}
Our implementation supports three observation presets derived from VisualWebArena~\citep{koh2024visualwebarena}: a text-only accessibility tree, a multimodal accessibility tree paired with the current viewport screenshot, and a set-of-marks representation that overlays numbered element identifiers on the screenshot.
All reported experiments use the multimodal accessibility-tree preset.
At each step, the agent receives the current task instruction, page URL, previous action, browser accessibility tree with element identifiers, and a screenshot of the current viewport.

\mypara{Action Space}
The action space follows the ID-based browser-action interface of VisualWebArena~\citep{koh2024visualwebarena} and contains 10 action types.
These actions support element interaction, text entry, keyboard input, page and container scrolling, within-site navigation, and episode termination.
Agents cannot create, switch, or close browser tabs, or navigate outside the website associated with the current task.
\Cref{tab:agent-action-space} provides the complete action grammar.

\begin{table}[h]
\centering
\caption{ID-based browser actions available to the target web agents.
Element IDs refer to the agent's current observation.}
\label{tab:agent-action-space}
\small
\setlength{\tabcolsep}{4pt}
\renewcommand{\arraystretch}{1.06}

\begin{tabularx}{\linewidth}{
@{}
>{\raggedright\arraybackslash}p{0.4\linewidth}
>{\raggedright\arraybackslash}X
@{}
}
\toprule
\textbf{Action}
& \textbf{Description} \\
\midrule

\texttt{click [id]}
& Click the element identified by \texttt{id}. \\

\texttt{clear [id]}
& Clear the contents of a text-input element. \\

\texttt{type [id] [text] [0|1]}
& Enter text into an element, optionally appending a newline. \\

\texttt{hover [id]}
& Move the pointer over an element. \\

\texttt{press [key\_comb]}
& Issue a keyboard key or key combination. \\

\texttt{scroll [up|down]}
& Scroll the current page in the specified direction. \\

\texttt{scroll [id] [up|down]}
& Scroll an identified element or nested container. \\

\texttt{goto [url]}
& Navigate to a URL within the current task website. \\

\texttt{go\_back} / \texttt{go\_forward}
& Move backward or forward through in-site browser history. \\

\texttt{stop [answer]}
& Terminate the episode and optionally return an answer. \\

\bottomrule
\end{tabularx}
\end{table}

\subsection{Seed-Sampling Details}
\label{app:seed-sampling}
During the adaptive stage, \method{} generates each new candidate by selecting an experience record and mutating its attack plan.
The selected record serves as the mutation seed, allowing previous execution outcomes to guide subsequent generation.
For each scheduler-assigned context, we first retrieve experience records associated with the same benign instruction and otherwise fall back to records from the same benign-task template.
Let $n$ denote the number of retrieved records and let ${\mathrm{ASR}}$ denote their running verifier-confirmed attack success rate.
When $n<8$ or ${\mathrm{ASR}}<0.5$, we use the low-evidence regime in \Cref{tab:seed-weights}, which favors verified successes as reliable starting points for mutation.
Otherwise, we use the refinement regime, which prioritizes near-misses to address the remaining failure stages.
Within the near-miss categories, benign-incomplete records receive greater weight because they provide stronger evidence that the attack diverted the agent from the authentic task.
Because every category has a positive weight, lower-priority records may still be sampled.

\begin{table}[h]
\centering
\caption{Relative weights for mutation-seed sampling under the low-evidence regime and refinement regime.}
\label{tab:seed-weights}
\small
\setlength{\tabcolsep}{5pt}
\renewcommand{\arraystretch}{1.08}

\begin{tabularx}{0.9\linewidth}{
@{}
>{\raggedright\arraybackslash}p{0.4\linewidth}
*{2}{>{\centering\arraybackslash}X}
@{}
}
\toprule

\multirow[c]{2}{*}{\textbf{Seed Category}}
& \textbf{Low-evidence}
& \textbf{Refinement} \\
& \textbf{Regime}
& \textbf{Regime} \\

\midrule

Success
& 5
& 2 \\

Near-miss, benign incomplete
& 4
& 5 \\

Near-miss, benign complete
& 3
& 4 \\

Blocked
& 2
& 3 \\

\bottomrule
\end{tabularx}
\end{table}

\subsection{Mutation Prompts}
\label{app:mutation-prompts}
For each adaptive proposal round, we construct a JSON-serialized prompt that combines fixed mutation instructions with rollout-specific context.
The fixed instructions define the permitted modifications, output format, and validity constraints.
The rollout-specific context contains the parent attack plan, benign user goal, outcome category, verifier results, attack-funnel signals, delivery configuration, and site-specific catalog constraints.
The program additionally assigns $k$ objective slots, each specifying a catalog-provided malicious objective, its parameters, action URL, attacker instruction, and a unique slot identifier.
The proposal model must yield one mutation per slot, preserving its assigned objective or parameters.

The mutation router determines which aspects of the parent plan may be revised according to its attack type and observed failure stage.
For user-level attacks, only the injected instruction may be revised.
For environment-level attacks, the prompt emphasizes placement and timing when the injection was not observed, CTA wording and visual salience when it was observed but not engaged, and continuation instructions when the malicious path was reached without completing the objective.
For successful seeds, the program may assign new catalog-backed objectives, while the proposal model varies only their expression and delivery.

The proposal model must return a single JSON object containing exactly $k$ mutations with distinct slot identifiers.
Each mutation must preserve the catalog-provided attacker instruction and, when applicable, its action URL and CTA label.
The model may not introduce new objective parameters, entity identifiers, routes, verifier rules, rendering logic, HTML, or arbitrary CSS.

\subsection{Mutation Output Schema}
\label{app:mutation-output-schema}
The proposal model does not directly generate executable attack plans.
Instead, it returns candidate revisions through a constrained JSON interface, allowing each revision to be bound to a program-assigned objective and validated before execution.
The output is a single JSON object containing a \texttt{mutations} array.
Each entry must bind to a unique assigned objective slot through \texttt{slot\_id} and include at least one permitted update listed in \Cref{tab:mutation-schema}. 
No other fields are accepted.

The system normalizes the returned mutations and compiles them into executable attack plans before candidate selection.
It removes entries with missing or duplicate slot identifiers, unsupported fields, invalid values, or payloads that omit required catalog-provided instructions, URLs, or CTA labels.
It also rejects duplicate plans and candidates that cannot be instantiated by the site-specific compiler.
If fewer than the required number of valid mutations remain, the unfilled slots receive catalog candidates that pass the same compiler validation.

\begin{table}[h]
\centering
\caption{Editable fields in the mutation-output schema.}
\label{tab:mutation-schema}
\small
\setlength{\tabcolsep}{4pt}
\renewcommand{\arraystretch}{1.08}

\begin{tabularx}{\linewidth}{
@{}
>{\raggedright\arraybackslash}p{0.27\linewidth}
>{\raggedright\arraybackslash}X
@{}
}
\toprule
\textbf{Field group}
& \textbf{Allowed output} \\
\midrule

Slot binding
& \texttt{slot\_id}, which must identify one assigned and previously unused objective slot. \\

Initial payload
& \texttt{payload\_override}, which revises the initial injected instruction or surface copy. \\

Delivery
& \texttt{delivery\_param\_overrides} and \texttt{delivery\_parameter\_id}, restricted to catalog-supported delivery fields. \\

Display timing
& \texttt{show\_after\_agent\_steps}, restricted to an integer from 1 to 3. \\

Surface style
& \texttt{surface\_style\_overrides}, restricted to environment-level attacks and whitelisted color, font-size, and font-family values. \\

Continuation
& \texttt{phase2\_payload\_override} and the corresponding continuation-phase delivery and style fields. \\

\bottomrule
\end{tabularx}
\end{table}

\subsection{Attack-Type Breakdown}
\label{app:attack-type-breakdown}
The main evaluation reports performance by website and target backbone without separating rollouts by attack type.
To complement this view, we regroup the evaluation results by attack type and target backbone.
\Cref{tab:attack-type-by-model} reports the ASR and CTA for each group and CTA is reported only for environment-level attacks.

Across all rollouts, direct prompt injection achieves the highest ASR of 31.5\%.
Among environment-level attacks, distraction and pop-up injection obtain ASRs
of 17.8\% and 17.1\%, respectively.
Pop-up injection achieves the highest CTA rate of 52.4\%, followed by
distraction injection at 47.0\%.
The difference between CTA and ASR indicates that initial engagement with
injected content frequently does not translate into successful completion of the malicious objective.
Performance also varies substantially across backbones, showing that
vulnerability to different delivery mechanisms is strongly model dependent.

\begin{table}[h]
\centering
\caption{
Attack-type performance of \method{} by target backbone model.
Each cell reports ASR/CTA (\%).
}
\label{tab:attack-type-by-model}
\small
\setlength{\tabcolsep}{0.8pt}
\renewcommand{\arraystretch}{1.12}

\begin{tabularx}{\columnwidth}{
@{}
>{\raggedright\arraybackslash}p{0.24\columnwidth}
>{\centering\arraybackslash}p{0.115\columnwidth}
>{\centering\arraybackslash}p{0.138\columnwidth}
*{3}{>{\centering\arraybackslash}X}
@{}
}
\toprule
\raisebox{1ex}{\textbf{Model}}
& \shortstack{\textbf{Direct}\\\textbf{PI}}
& \shortstack{\textbf{Indirect}\\\textbf{PI}}
& \shortstack{\textbf{Pop-up}\\\textbf{Injection}}
& \shortstack{\textbf{Distraction}\\\textbf{Injection}}
& \shortstack{\textbf{Ad}\\\textbf{Injection}} \\
\midrule

GPT-4o
& 56.6/--
& 38.9/46.3
& 42.3/83.3
& 39.3/89.3
& 29.3/56.3 \\

GPT-6.1 Sol
& 32.5/--
& 0.0/0.0
& 0.5/5.4
& 0.0/2.5
& 0.0/3.9 \\

Claude Sonnet 4.5
& 22.5/--
& 9.8/29.4
& 14.0/75.8
& 24.5/80.6
& 8.3/58.0 \\

Claude Sonnet 5.5
& 39.8/--
& 2.2/0.0
& 2.3/2.3
& 0.0/1.9
& 0.0/1.8 \\

Qwen3.7-Flash
& 23.6/--
& 6.3/14.6
& 18.6/69.7
& 13.2/42.6
& 14.1/34.8 \\

Qwen3-VL
& 35.5/--
& 0.0/0.0
& 8.0/69.2
& 9.1/40.1
& 1.7/15.0 \\

GLM-4.6V-Flash
& 15.4/--
& 15.4/69.2
& 8.4/45.6
& 7.4/49.5
& 6.6/54.3 \\

GLM-5V-Turbo
& 38.8/--
& 29.6/63.0
& 37.9/63.1
& 37.4/59.9
& 38.1/55.9 \\

\midrule

\textbf{Aggregate}
& \textbf{31.5/--}
& \textbf{15.9/25.8}
& \textbf{17.1/52.4}
& \textbf{17.8/47.0}
& \textbf{13.9/34.6} \\

\bottomrule
\end{tabularx}
\end{table}

\subsection{Runtime Analysis}
\label{app:runtime-analysis}

\begin{table}[h]
\centering
\caption{
Average wall-clock execution time per completed valid rollout across websites and victim backbones.
Values are reported in seconds.
}
\label{tab:runtime-per-rollout}
\small
\setlength{\tabcolsep}{2pt}
\renewcommand{\arraystretch}{1.10}
\renewcommand{\tabularxcolumn}[1]{m{#1}}

\begin{tabularx}{\columnwidth}{
@{}
>{\raggedright\arraybackslash}p{0.25\columnwidth}
*{4}{>{\centering\arraybackslash}X}
@{}
}
\toprule
\textbf{Website}
& \textbf{GPT-4o}
& \shortstack[c]{\textbf{GPT-6.1}\\\textbf{Sol}}
& \shortstack[c]{\textbf{Claude}\\\textbf{Sonnet 4.5}}
& \shortstack[c]{\textbf{Claude}\\\textbf{Sonnet 5.5}} \\
\midrule
Coursera   & 51.6  & 557.3 & 123.6 & 85.1  \\
Facebook   & 102.0 & 72.5  & 148.0 & 156.6 \\
GitHub     & 97.0  & 100.6 & 110.1 & 149.4 \\
Headspace  & 95.4  & 89.3  & 130.3 & 124.9 \\
JD         & 126.3 & 100.8 & 158.5 & 130.6 \\
Skyscanner & 98.4  & 135.4 & 95.8  & 102.9 \\
Slack      & 95.5  & 184.9 & 128.7 & 268.7 \\
Walmart    & 94.0  & 541.2 & 128.3 & 139.1 \\
YouTube    & 88.8  & 191.3 & 153.7 & 84.7  \\
\midrule
\textbf{Avg.}
& \textbf{94.3}
& \textbf{219.3}
& \textbf{130.8}
& \textbf{138.0} \\
\bottomrule
\end{tabularx}

\vspace{6pt}

\begin{tabularx}{\columnwidth}{
@{}
>{\raggedright\arraybackslash}p{0.25\columnwidth}
*{4}{>{\centering\arraybackslash}X}
@{}
}
\toprule
\textbf{Website}
& \textbf{Qwen3-VL}
& \shortstack[c]{\textbf{Qwen3.7}\\\textbf{Flash}}
& \shortstack[c]{\textbf{GLM-4.6V}\\\textbf{Flash}}
& \shortstack[c]{\textbf{GLM-5V}\\\textbf{Turbo}} \\
\midrule
Coursera   & 436.8 & 61.0 & 98.2  & 68.2  \\
Facebook   & 238.6 & 66.9 & 135.8 & 97.0  \\
GitHub     & 252.6 & 57.8 & 128.1 & 81.7  \\
Headspace  & 242.2 & 88.1 & 183.6 & 87.5  \\
JD         & 221.9 & 63.6 & 168.2 & 109.9 \\
Skyscanner & 283.1 & 54.1 & 132.2 & 143.8 \\
Slack      & 536.1 & 65.9 & 191.9 & 103.6 \\
Walmart    & 254.5 & 59.3 & 119.5 & 205.9 \\
YouTube    & 231.3 & 58.4 & 127.4 & 85.1  \\
\midrule
\textbf{Avg.}
& \textbf{299.7}
& \textbf{63.9}
& \textbf{142.8}
& \textbf{109.2} \\
\bottomrule
\end{tabularx}
\end{table}

We report the wall-clock execution time of \method{} across all 8 victim backbones and 9 websites.
For each valid formal rollout, we measure the elapsed time from the creation of its execution artifact to the completion of the final trajectory.
The measurement therefore includes victim-model inference, browser interaction, environment execution, and verifier evaluation.
It excludes candidate generation before rollout dispatch, queueing between rollouts, interrupted executions, and infrastructure failures that were not committed as formal experience records.
\Cref{tab:runtime-per-rollout} reports the mean execution time per valid rollout.

The results show that the runtime of \method{} depends primarily on the latency of the victim backbone and the number of interaction steps required by each website.
Qwen3.7-Flash has the lowest average execution time at 63.9 seconds per rollout, followed by GPT-4o at 94.3 seconds and GLM-5V-Turbo at 109.2 seconds.
Qwen3-VL requires the longest average runtime at 299.7 seconds per rollout, with particularly long executions on Coursera and Slack.
GPT-6.1 Sol also exhibits substantial site-dependent variation, primarily due to long executions on Coursera and Walmart.
Because the victim models are served through a mix of local GPUs and remote APIs, these measurements represent practical rollout latency under our evaluation setup rather than controlled model-serving throughput.

\end{document}